\documentclass[twocolumn]{openjournal}
\usepackage{graphicx}   % Including figure files
\usepackage{amsmath}    % Advanced maths commands
\usepackage{amssymb,amsfonts}    % Extra maths symbols
\usepackage{orcidlink}
\usepackage{verbatim}
\usepackage{hyperref}
\usepackage{xcolor}
\usepackage{longtable}
\usepackage{booktabs}

\shorttitle{Theta Eridani: a millenary stellar transient}
\shortauthors{Idel Waisberg \& Boaz Katz}

\begin{document}

\title{The forgotten bright star:\\Theta Eridani as a millenary stellar transient\\observed by Hipparchus, Ptolemy and al-Sufi}

\footnotetext[]{Based on observations collected at the European Southern Observatory, Chile, Program IDs 074.A-9002(A), 099.D-2031(A), 114.27DS.001}

\newcommand{\weizmann}{Department of Particle Physics and Astrophysics, Weizmann Institute of Science, Rehovot 76100, Israel}

\email{email: idelwaisberg@gmail.com}

\author{\vspace{-1.2cm}Idel Waisberg\,\orcidlink{0000-0003-0304-743X}$^{1}$ \& Boaz Katz\orcidlink{0000-0003-0584-2920}$^{2}$}

\affiliation{$^1$Independent researcher, Lambda Ophiuchi Ltda}
\affiliation{$^2$\weizmann}

\begin{abstract}
Theta Eridani is a V=2.9 star that was nonetheless reported as one of the thirteen brightest stars in the night sky by both Ptolemy in his \textit{Almagest} (137 AD) and by al-Sufi in his \textit{The Book of Fixed Stars} (964 AD), in addition to being previously referred by Hipparchus (129 BC) as a particularly bright star. The discrepancy between its historical and modern visual magnitude $\Delta V \sim 2.7$ is the highest among the $\sim 1000$ stars in the \textit{Almagest}. Theta Eridani is actually a triple star system, and here we combine interferometric data from VLTI/PIONIER and VLTI/GRAVITY, spectroscopic data from ESPaDOns and FEROS, and photometric data from TESS in order to solve for the orbital parameters, masses and radii of the close inner binary Theta Eridani Aa+Ab. We find that it is a tight eccentric binary ($a=0.083 \text{ au}$, $e=0.105$) of intermediate-mass stars ($M_{Aa}\simeq 2.3 M_{\odot}$, $M_{Ab}\simeq 2.2 M_{\odot}$) that are extended to $\sim 80\%$ of their Roche lobe radii ($R_{Aa}\simeq 4.3 R_{\odot}, R_{Ab} \simeq 4.0 R_{\odot}$), resulting in prominent ellipsoidal oscillations in the lightcurve. We also find that the primary is in a very special phase of its evolution in which it has just finished core hydrogen burning. The remarkable combination of orbital and stellar parameters hints that the historical brightening of Theta Eridani was due to a millenary transient phase powered by orbital energy extraction during a long-lived ``common envelope'' stage triggered by eccentric Roche lobe overflow in a previously more eccentric binary ($e\simeq0.6$). This strengthens the case that the apparent brightening was real and not due to an error by three different ancient observers, as has been commonly claimed in the past. 
\end{abstract}

% https://astrothesaurus.org

\section{Introduction}
\label{sec:introduction}

\begin{figure*}[]
\includegraphics[width=2\columnwidth]{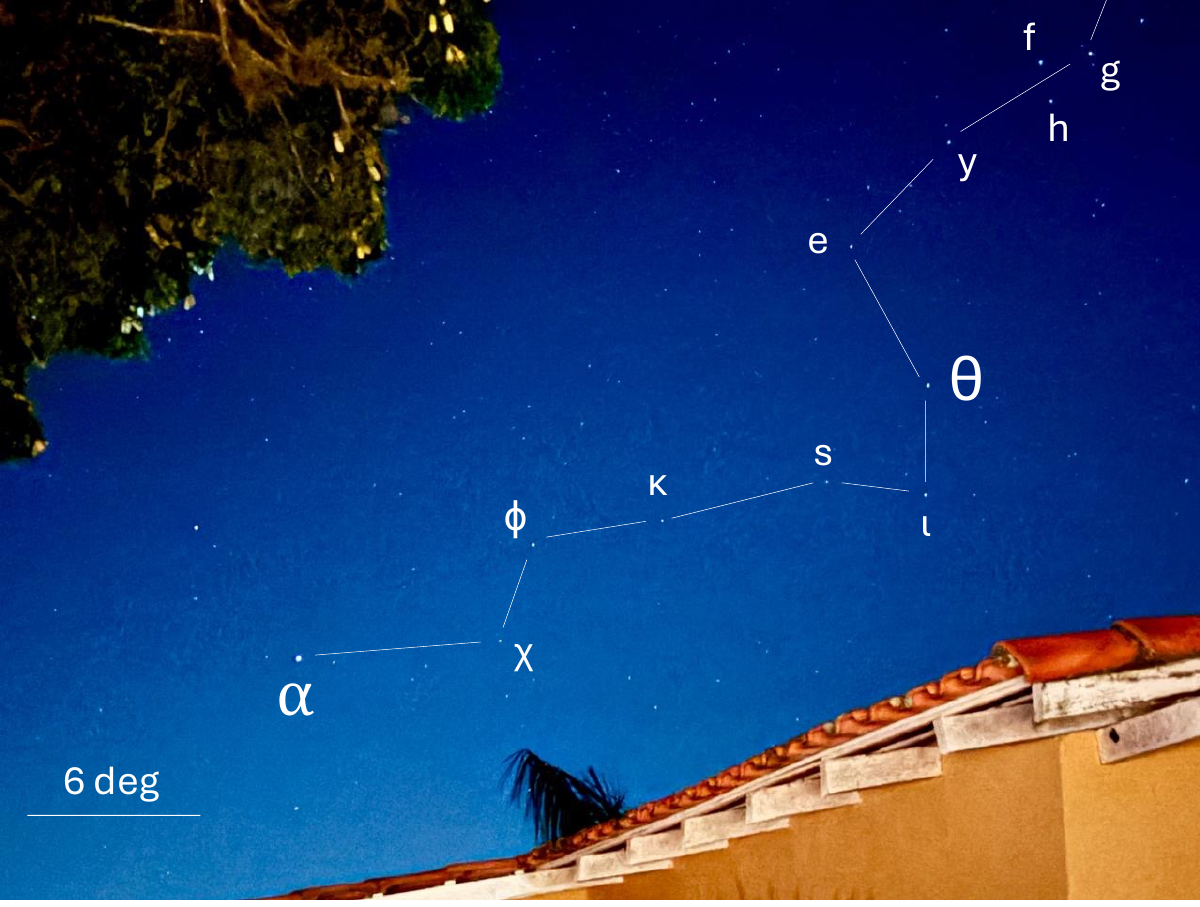}
\caption{Smartphone photo of the southern part of Eridanus seen from Jacutinga, MG, Brazil on 2026-03-22 UTC 23:06.}
\label{fig:sky_image}
\end{figure*}

Theta Eridani ($\theta$ Eri) is a V=2.9 star that used the demark the end of the Eridanus constellation, hence its proper name \textit{Acamar} (``the last of the River''). With the first European voyages to the Southern Hemisphere, the Eridanus constellation was extended further south starting with Petrus Plancius’s celestial globe of 1598, causing $\theta$ Eri to be supplanted by the (currently) much brighter and virtually homonymous (V=0.45) Alpha Eridani ($\alpha$ Eri) = \textit{Achernar}, which was simply not visible to Greek and Arab astronomers due to a combination of its souther declination and Earth's precession. Figure \ref{fig:sky_image} shows a smartphone photo of the southern part of Eridanus containing both $\theta$ and $\alpha$ Eri as seen from Brazil. 

Albeit an apparently unassuming star, there is an unsolved mystery regarding Theta Eridani: it was listed in both Ptolemy's \textit{Almagest} as well as in al-Sufi's \textit{The Book of Fixed Stars} as one of the thirteen brightest stars in the night sky, in stark contrast to its modern and relatively humble brightness. Whether this discrepancy is real or the result of misjudgement by the ancient observers was a topic of debate around a century ago \citep{Anderson1893,Gore1900,Webb1916,Backhouse1916} to which there has never been a convincing resolution. 

Theta Eridani is actually not one but rather three stars. It consists of the $\rho=8.3" \leftrightarrow 425 \text{ au}$ visual binary $\theta$ Eri A (HD 18622, HR 897) and $\theta$ Eri B (HD 18623, HR 898), wherein the visual primary $\theta$ Eri A is also a very close binary. It had been previously reported as SB2 \citep{Wright1905,Faraggiana2001} and we have recently reported a VLTI/GRAVITY interferometric observation in which the close binary was spatially resolved in the near-infrared K band while the visual secondary was confirmed to be a single star \citep{Waisberg2025}. Through isochrone fitting we estimated masses $M_{Aa} \simeq 2.39 M_{\odot}$, $M_{Ab} \simeq 2.35 M_{\odot}$ and $M_{B} \simeq 2.30 M_{\odot}$. The close binary nature of $\theta$ Eri A has also been confirmed by its lightcurve from the Transiting Exoplanet Survey Satellite \citep[TESS;][]{Ricker2015}, which shows ellipsoidal variations of amplitude $\Delta m \simeq 0.02$ with a period $P=4.1077$ days \citep[e.g.][]{Watson2006}.

The goal of this paper is to further characterize the system and explore whether there is a reasonable explanation for the historical brightness discrepancy. Specifically, we make use of archival interferometric, spectroscopic and photometric observations to solve for the orbital parameters, dynamical masses and radii of the close binary $\theta$ Eri Aa+Ab. 

This paper is organized as follows. In Section \ref{sec:history}, we discourse on the history of Theta Eridani and explain why in our view there is no strong reason to doubt the accuracy of the ancient reports. In Section \ref{sec:observations}, we describe the interferometric, spectroscopic and photometric observations. Section \ref{sec:results} contains the results, including the orbital fitting, lightcurve, photometric and spectral analyses. In Section \ref{sec:discussion}, we discuss a possible mechanism to explain the brightening episode and the possibility that there are more such transients in modern photometric data. We conclude this paper in Section \ref{sec:conclusion}. 

\section{The story of Theta Eridani} 
\label{sec:history}

This section is named after the \textit{Knowledge} article ``The story of $\theta$ Eridani'' \citep{Anderson1893}, in which a most intriguing account of the issue can be found. Here we present further insight into it based on our own research and relevant later works. Table \ref{table:historical_catalogs} summarizes important information concerning Theta Eridani for the different observers that will be discussed, including their place and time of observation, the declination of $\theta$ Eri at the time taking into account Earth's precession, the difference between the reported ecliptic longitude ($\lambda$) and latitude ($\beta$) and the ``true'' one based on data from the Hipparcos mission and an accurate equinox precession model\footnote{The positional error in the table already corrects for Ptolemy's systematic offset of $1^{\circ}$ in ecliptic longitude.} \citep{Verbunt2012}, the maximum altitude reached by $\theta$ Eri and the corresponding airmass \citep{Kasten1989} and atmospheric visual extinction $k_v$  using a typical visual extinction coefficient $0.2 \text{ mag}\text{ airmass}^{-1}$. 

\begin{table}
\centering
\caption{\label{table:historical_catalogs} Historical catalog properties concerning Theta Eridani.}
\begin{tabular}{ccccc}
\hline \hline
& Hipparcos & Ptolemy & al-Sufi & Ulugh Beg \\ [0.3cm]

\shortstack{Place\\Latitude($^{\degr}$)} & \shortstack{Rhodes\\+36.4} & \shortstack{Alexandria\\+31.2} & \shortstack{Shiraz\\+29.6} & \shortstack{Samarkand\\+39.7} \\ [0.3cm]

Time & 129 BC & 137 AD & 964 AD & 1437 AD \\ [0.3cm]

\shortstack{Declination\\(deg)} & $-49.8$ & $-48.7$ & $-46.0$ & $-44.1$ \\ [0.3cm]

\shortstack{$(\Delta \lambda$,$\Delta \beta)$ (') \\ $\rho$ (')} & - & \shortstack{(-247.5,-22.8)\\148.3} & - & \shortstack{(-19.3,-1.9)\\11.6}  \\ [0.3cm]

\shortstack{max. altitude \\ airmass} & \shortstack{3.8\\12.9} & \shortstack{10.1\\5.5} & \shortstack{14.4\\4.0} & \shortstack{6.2 \\ 8.6} \\ [0.3cm] 

$k_V$ & 2.6 & 1.1 & 0.8 & 1.7 \\ [0.3cm]

\hline
\end{tabular}
\end{table}

Theta Eridani was first mentioned by Hipparchus (Rhodes, 129 BC) in his surviving work \textit{Commentary on Aratus} as ``the brightest and preceding and southernmost of all in the River''. The coordinates provided by Hipparchus allow to definitely exclude the possibility that he was referring to $\alpha$ Eri, which is much more southern and was actually invisible to Hipparchus. $\theta$ Eri itself achieved a maximum altitude of only $3.8^{\circ}$ at Rhodes in 129 BC and would be subject to significant atmospheric extinction of about 2.6 mag. Therefore, even though it is currently the third visually brightest star in Eridanus even at its current $V=2.9$ magnitude (after $\alpha$ Eri and $\beta$ Eri, the latter of which is only slightly brighter at $V=2.8$ and is 35 deg distant near the Eridanus riverhead next to Orion), it is difficult to imagine why Hipparchus would be so adamant about its brightness considering that there are many stars near $\theta$ Eri that are just slightly ($\Delta V \sim 0.5$) fainter (as can be seen in Figure \ref{fig:sky_image}), unless $\theta$ Eri were indeed significantly brighter at his time. 

The strongest piece of evidence for $\theta$ Eri being brighter in the past comes from the \textit{Almagest} by Claudius Ptolemy \citep[Alexandria, 137 AD; see][for an English translation]{Toomer1998}. The catalog contains 1,025 distinct stars and each received a numerical magnitude in integer steps between 1 and 6, with a possible qualifier for 'bright' (b) or 'faint' (f) to capture intermediate values\footnote{There are also a small number of stars classified as faint (12 stars) or nebulous (5 stars) that do not have an assigned magnitude.}. $\theta$ Eri is one of only fifteen stars that received the smallest magnitude of 1. Two of these stars (namely, \textit{Betelgeuse} and \textit{Denebola}) have a qualifier 'f' while none has a qualifier 'b'. Therefore, $\theta$ Eri is among the thirteen brightest stars in the \textit{Almagest}. 

\begin{figure*}[]
\centering
\includegraphics[width=\textwidth]{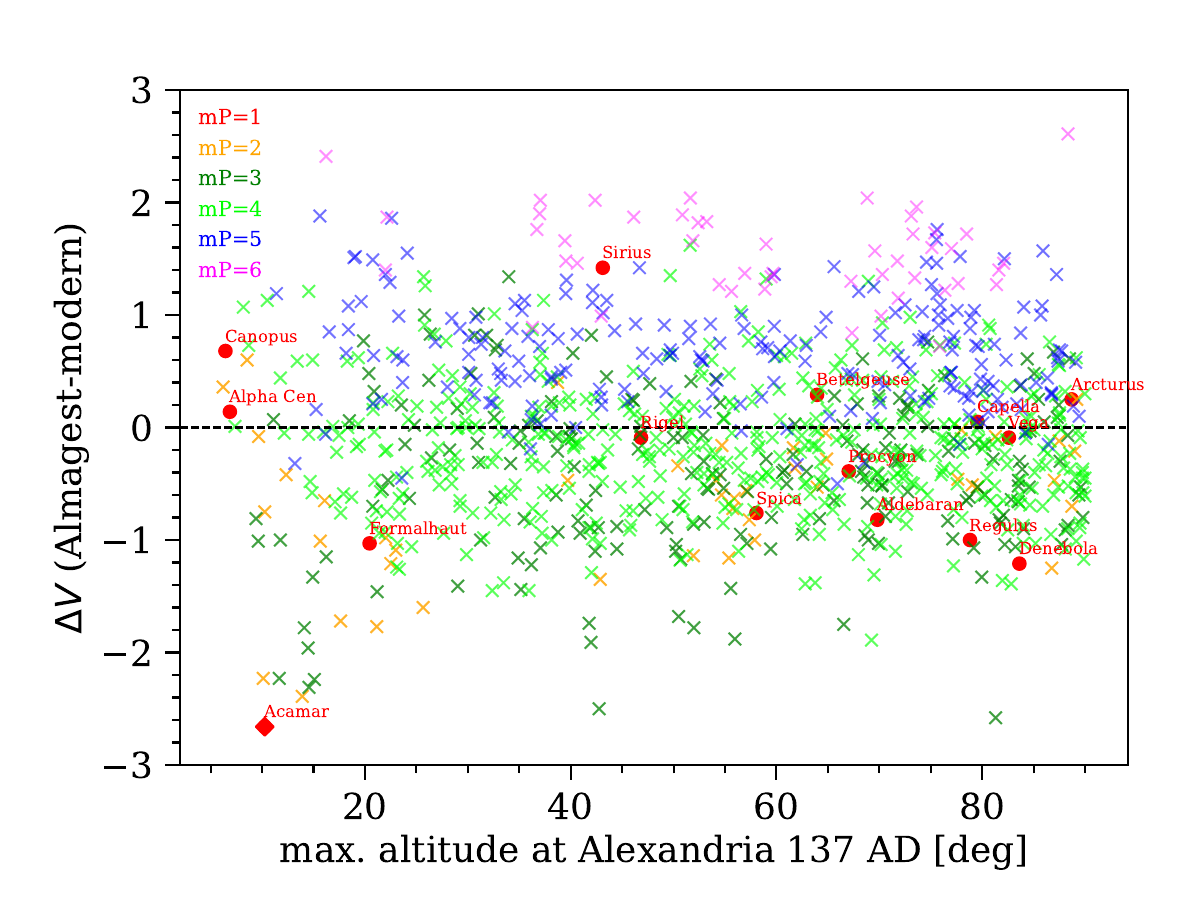}
\caption{\label{fig:historical} Visual magnitude discrepancy between the \textit{Almagest} catalog and modern observations for 992 stars as estimated in \cite{Protte2020} as a function of their culmination altitude at Alexandria in 137 AD. The stars are colored by their \textit{Almagest} magnitude (mP) and the fifteen magnitude 1 stars are labeled. Theta Eridani=\textit{Acamar} has the largest deviation among all stars.}
\end{figure*}

\cite{Verbunt2012} correlated the stars in the \textit{Almagest} with ESA's Hipparcos mission catalog, finding high fidelity identifications for the vast majority of stars. \cite{Protte2020} then performed a detailed statistical analysis of the photometric properties of the \textit{Almagest} based on a sample of 992 stars from the \cite{Verbunt2012} catalog\footnote{The 33 excluded stars include the 17 ``faint" or ``nebulous" stars with unassigned magnitudes as well as stars with an identification flag of only ``probable" or below, which are mostly faint stars, namely one star with magnitude 3, six stars with magnitude 4, seven stars with magnitude 5 and one star with magnitude 6.}. \cite{Protte2020} then devised a linear conversion of the \textit{Almagest} magnitude to the Johnson V magnitude as well as (usually minor) correction terms for color dependence, atmospheric extinction and level of background light. The \textit{Almagest} magnitude of $\theta$ Eri converts to a Johnson magnitude of 0.22 (0.11 without the correction terms), compared to its modern V magnitude of 2.88. The discrepancy $\Delta V \simeq 2.7$ is not only the highest among the magnitude 1 stars in the \textit{Almagest} but also among \textit{all} 992 stars. Figure \ref{fig:historical} shows the magnitude discrepancy between the \textit{Almagest} and modern observations for the 992 stars as a function of their culmination altitude at Alexandria in 137 AD \footnote{Note that in the tables in \cite{Protte2020} HIP 71681 (Alpha Centauri) is given a modern magnitude $V=1.35$ corresponding to the fainter component $\alpha$ Cen B, when in reality the combined magnitude of the system $\alpha$ Cen A+B ($V=-0.27$) should be used. This has been taken into account in Figure \ref{fig:historical}.}. The stars are colored by their \textit{Almagest} magnitude and the fifteen magnitude 1 stars are labeled. 

It is widely accepted that Ptolemy's \textit{Almagest} is at least partly based on Hipparchus' lost star catalog; the strongest evidence for this claim is that the stellar positions in the \textit{Almagest} have a clear and characteristic systematic error of $1^{\degr}$ in longitude relative to their expected positions at the time, which is consistent with Ptolemy's slightly wrong value for Earth's precession rate applied to Hipparchus' epoch \citep[e.g.][]{Duke2003,BaigetOrts2026}. Therefore, one may wonder whether it is possible that Ptolemy did not independently observe Theta Eridani. However, there is no doubt that he did: not only was $\theta$ Eri rather easily observed from Alexandria in 137 AD (with a culmination altitude of $10.1^{\circ}$) but it is also explicitly mentioned as a first magnitude star in his other work \textit{Phaseis} (Calendar), in which its rising and setting times are given together with those of the other fourteen stars of magnitude 1 \citep{Anderson1893,Webb1916}. It is interesting to note, however, that the ecliptic longitude of $\theta$ Eri reported in the \textit{Almagest} is off by about $\Delta \lambda \sim 4^{\circ}$ relative to its expected position (already correcting for the $1^{\circ}$ systematic error). While at the high end of the positional errors, there are many stars with similar and even larger positional errors so that the relevance of the offset is inconclusive. 

\begin{figure*}[]
\centering
\includegraphics[width=\textwidth]{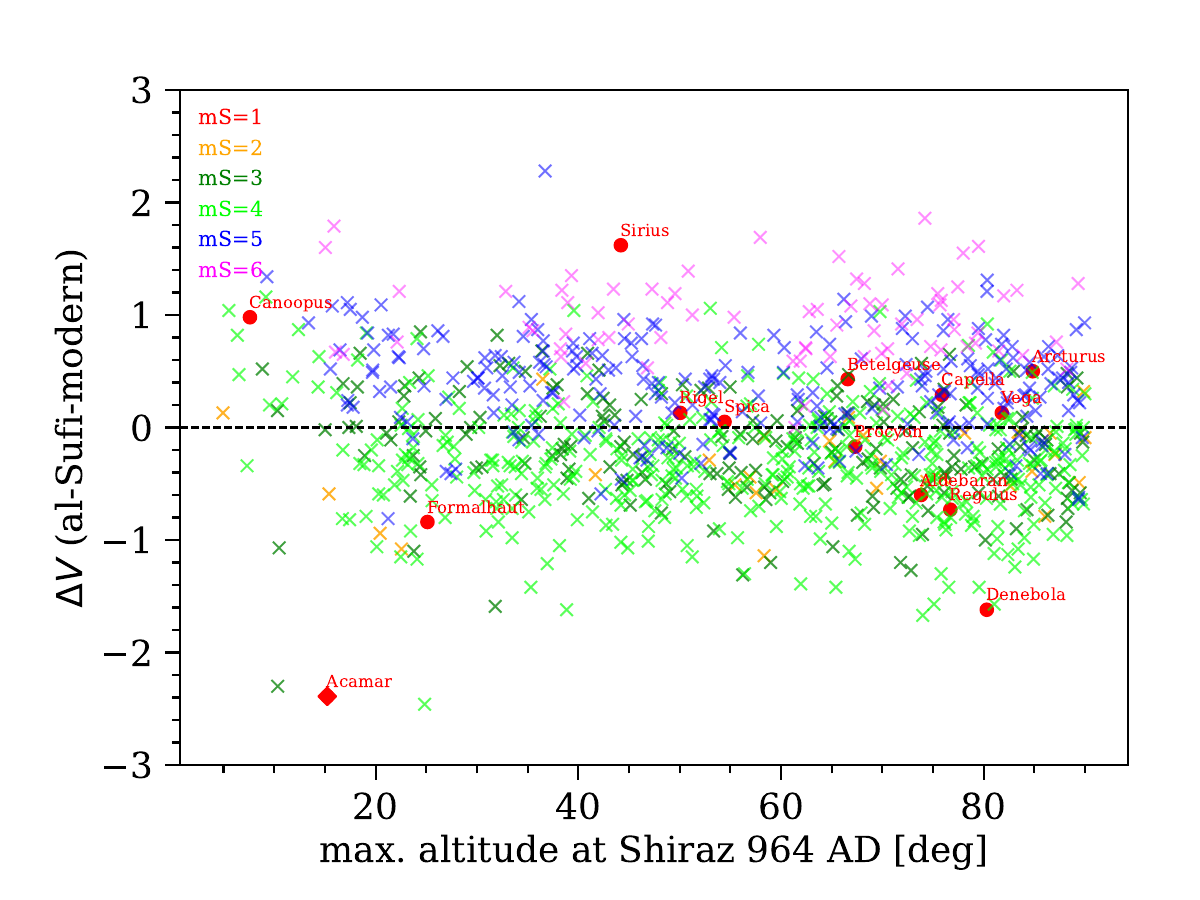}
\caption{\label{fig:historical_al-sufi} Visual magnitude discrepancy between al-Sufi's \textit{The Book of Fixed Stars} and modern observations for 990 stars as estimated in \cite{Protte2020} as a function of their culmination altitude at Shiraz in 964 AD. The stars are colored by their al-Sufi magnitude (mS) and the fourteen magnitude 1 stars are labeled.}
\end{figure*}

We next find Theta Eridani in \textit{The Book of Fixed Stars} by Abd al-Rahman al-Sufi \citep[Shiraz, 964 AD; see][for French and English translations]{Schjellerup1874,Hafez2010}, where it is still listed as having magnitude 1. al-Sufi's catalog is heavily based on the \textit{Almagest}: it contains mostly the same stars (with some additions) and the positions correspond exactly to those in the Almagest with an addition of 12 deg 42 arcmin to the ecliptic longitude to account for the equinox precession. However, al-Sufi did revise many of the magnitudes and in most cases this led to a better agreement with modern values \citep{Verbunt2012}. Very illustrative examples are the two mP=2 stars that can be seen near $\theta$ Eri in Figure \ref{fig:historical}, namely HIP 38827 ($\chi$ Car) and HIP 95241 ($\beta$ Sgr): al-Sufi revised both of their magnitudes from 2 to 4 towards agreement with their modern values, and yet did not revise the magnitude of $\theta$ Eri. Can we be sure that al-Sufi independently observed $\theta$ Eri? The answer is yes: it was easily observable from Shiraz in 964 AD (with a culmination altitude of $14.4^{\circ}$) and his catalog contains a detailed description of its position relative to the surrounding stars $\iota$, $e$ $f$, $g$ and $h$ Eridani (marked in Figure \ref{fig:sky_image}) as detailed in \cite{Anderson1893}.
The magnitude 1 stars in the \textit{The Book of Fixed Stars} are almost the same as in the \textit{Almagest}, with the omission of Alpha Centauri (which was not part of the catalog). Therefore, like in the \textit{Almagest} $\theta$ Eri presumably still appeared to al-Sufi as one of very brightest stars in the night sky. In the case of al-Sufi, the visual magnitude discrepancy with modern observations, $\Delta V \simeq 2.4$, is the second highest among the 990 stars \citep{Protte2020}. Figure \ref{fig:historical_al-sufi} shows the analog of Figure \ref{fig:historical} for al-Sufi's \textit{The Book of Fixed Stars} based on the catalog of \cite{Protte2020}. 

Almost five centuries later we find $\theta$ Eri in the \textit{Zij-i Sultani} star catalog by Ulugh Beg \citep[Samarkand, 1437 AD; see][for an English translation]{Knobel1917}. This catalog was based on al-Sufi's and had the opposite purpose to it: to revise the positions rather than the magnitudes of the stars. As a result, the reported magnitudes are exactly the same as in \textit{The Book of Fixed Stars} but nearly all the positions are revised, except for 27 stars that were too far south for Ulugh Beg to observe \citep{Verbunt2012}. There is no question that $\theta$ Eri was independently observed by Ulugh Beg (it culminated at $6.2^{\circ}$ at Samarkand in 1437 AD) as its revised position is in much better agreement with its expected position at the time (Table \ref{table:historical_catalogs}) compared to the \textit{Almagest}. However, because Ulugh Beg did not revise any of the star magnitudes it is not possible to make any strong conclusions regarding its brightness at the time. 

\begin{figure*}[]
\includegraphics[width=2\columnwidth]{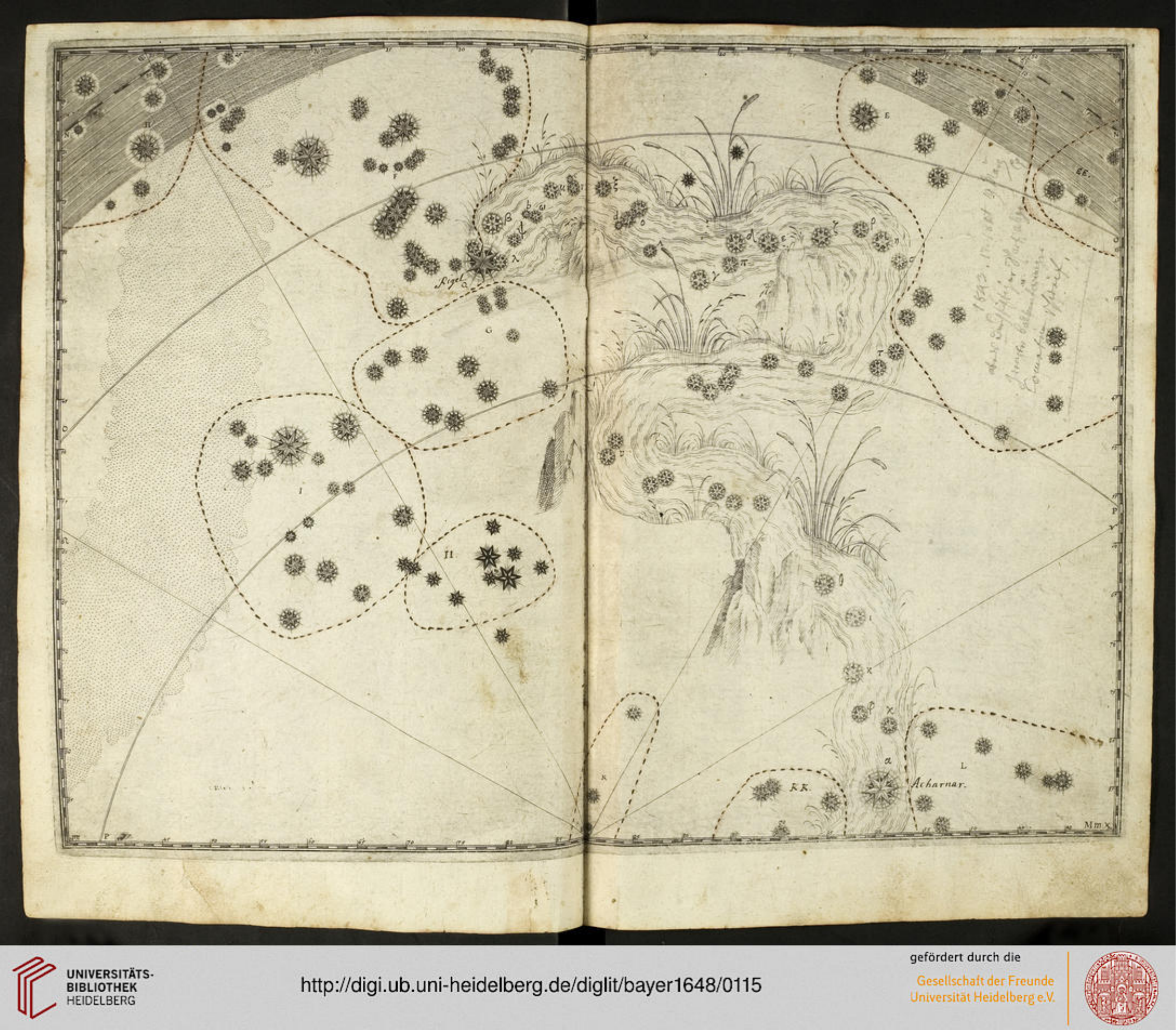}
\caption{Eridanus constellation in Johann Bayer's \textit{Uranometria} published in 1603. The extension of the constellation to \textit{Achernar}=$\alpha$ Eridani was based on the stellar globes containing the southern sky produced just a few years prior based on the \textit{Eerste Schipvaart} expedition. $\theta$ Eridani is already depicted as much fainter than $\alpha$.}
\label{fig:Uranometria}
\end{figure*}

$\theta$ Eri was again referenced about 150 years later in the southern sky globes, charts and catalogs created after the first European voyages to the Southern Hemisphere, specifically the first Dutch expedition to the East Indies from Amsterdam to Java and Sumatra (\textit{Eerste Schipvaart}). In Johann Bayer's \textit{Uranometria} published in 1603 \citep{Bayer1603}, the Eridanus constellation was already drawn extending down to $\alpha$ Eri, which is depicted as much brighter than $\theta$ Eri (Figure \ref{fig:Uranometria}). Bayer's extension of Eridanus (as well as his depiction of the southern constellations around the South Pole) were based on the stellar globes containing the southern sky produced just a few years prior. 

Also in 1603 Frederick de Houtman (one of the members of the Dutch expeditions) published a star catalog of the southern sky. de Houtman used a magnitude system analogous to Ptolemy using integer steps between 1 and 6, and assigned a magnitude 3 to $\theta$ Eri (and a magnitude of 1 to $\alpha$ Eri) that is consistent with its modern magnitude \citep{Knobel1917,Verbunt2011}. $\theta$ Eri was later featured in other southern star catalogs. In Edmond Halley's \textit{Catalogus Stellarum Australium} published in 1679 \citep{Halley1679} (and based on observations made in 1677-1678 in the island of St. Helena), $\theta$ Eri is reported with a magnitude of 3.0. In his catalog of fundamental stars published in 1757 \citep{Lequeux2014}, the French astronomer Nicolas-Louis La Caille also reported $\theta$ Eri as a third magnitude star based on observations made in 1751-52 in Cape Town. Therefore, there is no doubt that by around 1600 $\theta$ Eri had already settled to its modern third magnitude and has remained so until today. 

The historical records therefore suggest that Theta Eridani had a magnitude $V \sim 0.2$ for at least about a millennium (i.e. from Hipparchus in 129 BC to al-Sufi in 964 AD) and somewhere between the latter and 1597 it faded by almost three magnitudes to its modern $V=2.9$ brightness. This behavior is highly unusual and as far as we know unprecedented for a system of three A-type stars not far form the main-sequence. It is therefore important to address the alternative and more mundane explanations for its reported large brightness in the ancient times that have been previously put forward: 

\begin{enumerate}
\item \textit{Confusion with Alpha Eridani}: This is rather easily refuted. Not only was $\alpha$ Eri not visible to the ancient Greek and Arab astronomers, but the location of $\theta$ Eri was explicitly described both by Ptolemy in \textit{Phaseis} and by al-Sufi in the \textit{The Book of Fixed Stars}. Therefore the possibility that they learned about $\alpha$ Eri from travelers from more southern lands and were referring to it can also be excluded. 

\item \textit{A typographical error}: The possibility that a $\delta$ (4) was corrupted to an $\alpha$ (1) throughout the copies of the \textit{Almagest} is also easily refuted. $\theta$ Eri was explicitly mentioned as ``brilliant'' by Ptolemy and is part of the group of the fifteen brightest stars in his \textit{Phaseis} \citep{Webb1916}. It was also explicitly mentioned as first magnitude by al-Sufi, who did not revise its \textit{Almagest} magnitude even though he clearly observed it. 

\item \textit{Over-correction of atmospheric extinction}: The relatively low culminations of $\theta$ Eri for the ancient observers imply non-negligible amounts of visual atmospheric extinction, so it has been suggested that its reported brightness could have been a result of an overestimated extinction correction. This explanation is nonetheless not supported by the data. Analysis of the magnitude discrepancy between the \textit{Almagest} and modern observations as a function of culmination angle have shown that there is no evidence for atmospheric extinction residuals in the data \footnote{It should be noted that it is still somewhat of a mystery how Ptolemy achieved such precise extinction correction.} \citep{Schaefer2013,Protte2020}. In the case of al-Sufi, there appears to be a net residual atmospheric extinction term \citep[which is nonetheless much smaller than the expected extinction, implying that some form of atmospheric extinction correction was successfully applied;][]{Schaefer2013,Protte2020}, but it works on the opposite direction to explain $\theta$ Eri i.e. the extinction was slightly underestimated rather than over-corrected. Arguably the strongest argument that atmospheric extinction over-correction does explain the $\theta$ Eri discrepancy is that, as can be seen in Figure \ref{fig:historical}, two other southern mP=1 stars, \textit{Canopus} and Alpha Centauri, have lower culmination altitudes than $\theta$ Eri and yet their brightnesses were not overestimated in the ancient catalogs (\textit{Canopus} was rather underestimated). Therefore, while it is ultimately impossible to exclude the possibility that a specific and uncharacteristic misjudgement was made in the case of $\theta$ Eri, this does not find support in the data. 
\end{enumerate}

In conclusion, we find no convincing reason to refute the fact that Theta Eridani was visually brighter by about a factor of ten between about 2,000 to 1,000 yrs ago. The remaining of this paper is dedicated to constraining the orbital and stellar parameters of the system and presenting a possible mechanism to explain its rather extreme change in brightness.

\section{Observations}
\label{sec:observations}

\subsection{VLTI/PIONIER}
\label{obs:pionier}

$\theta$ Eri A was observed with the beam combiner PIONIER \citep{LeBouquin2011} coupled with the 1.8-meter Auxiliary Telescopes (ATs) during five epochs in 2017 (PI: Ireland). VLTI/PIONIER works in the near-infrared H band at low spectral resolution (in this case, with six spectral channels over the H band). In the first epoch, only three telescopes were used so the data consists of squared visibilities for three baselines and closure phases for one triangle. For the other four epochs, the four telescopes were used leading to six baselines and four closure triangles. In all cases, the individual exposure time is 0.5 ms and the observation block consists of five files with 51,200 exposures each. A summary of the observations, including the seeing, maximum projected baseline and corresponding angular resolution, are reported in Table \ref{table:pionier_obs}. 

\begin{table*}
\centering
\caption{\label{table:pionier_obs} Summary of archival VLTI/PIONIER observations of Theta Eridani A and best-fit binary model parameters.}
\begin{tabular}{cccccccccc}
\hline \hline
\shortstack{date\\MJD} & \shortstack{seeing\\@ 500 nm (")} & \shortstack{AT config.} & \shortstack{$B_{\mathrm{proj,max}}$ (m) \\$\theta_{\mathrm{max}}$ (mas)} & calibrator(s) & \shortstack{$\frac{f_{\mathrm{A_b}}}{f_{\mathrm{A_a}}}$\\H band\\(\%)} & \shortstack{$\Delta \alpha_*$\\(mas)} & \shortstack{$\Delta \delta$\\(mas)} \\ [0.3cm]

\shortstack{2017-07-16\\57950.420} & 1.1 & A1-G1-J3 & \shortstack{132.4\\2.4} & \shortstack{HD 10019\\HR 537} & 90.7 & $1.19\pm0.04$ & $-0.22\pm0.04$ \\ [0.3cm]

\shortstack{2017-07-25\\57959.392} & 1.0 & A1-G1-J2-K0 & \shortstack{123.4\\2.6} & HR 537 & 87.8 & $0.17\pm0.04$ & $-1.50\pm0.04$ \\ [0.3cm]

\shortstack{2017-07-27\\57961.438} & 0.5 & A1-G1-J2-J3 & \shortstack{132.0\\2.4} & HR 537 & 88.0 & $-0.42\pm0.04$ & $1.14\pm0.04$ \\ [0.3cm]

\shortstack{2017-07-28\\57962.432} & 0.9 & A1-G1-J2-J3 & \shortstack{132.1\\2.4} & HR 537 & 88.1 & $1.27\pm0.04$ & $0.51\pm0.04$ \\ [0.3cm]

\shortstack{2017-07-27\\57992.362} & 1.2 & A1-G1-J2-J3 & \shortstack{131.7\\2.4} & HD 17206 & 88.3 & $0.016\pm0.04$ & $-1.63\pm0.04$ \\ [0.3cm]

\hline
\end{tabular}
\end{table*}

In all cases $\theta$ Eri A was actually observed as a calibrator star. Fortunately, other calibrators were also observed in the same sequence, which allowed us to treat $\theta$ Eri A as a science target. The calibrator stars we used are noted in Table \ref{table:pionier_obs} and include HD 10019 (G8III, angular diameter $\theta=0.57 \text{ mas}$), HR 537 (K0III, $\theta=0.81 \text{ mas}$) and HD 17206 (F7V, $\theta=0.89 \text{ mas}$). 

We downloaded the data from the ESO archive, including the calibration files (dark and kappa matrix frames), and reduced the data using the default settings in the PIONIER data reduction software \texttt{pndrs} v3.94 \citep{LeBouquin2011}. The only additional step needed was to change the label of the observations of $\theta$ Eri A from calibrator to science for the final calibration step.

\subsection{ESPaDOns}

We downloaded four ``ready-to-use'' archival spectra of $\theta$ Eri A from PolarBase \citep{Petit2014}. The spectra were taken in 2014 and 2015 with the ESPaDOns spectropolarimeter \citep{Donati2006} at the Canada-France Hawaii Telescope. They cover the wavelength range 3700-10400{\AA} at a spectral resolution $R \approx 65,000$. There were several observations per epoch with individual exposure times of 40 or 45 s. We averaged the observations weighted by their SNR in order to end up with one spectrum per epoch. We note that the reduced spectra are already corrected to the heliocentric velocity. Details of the spectral observations can be found in Table \ref{table:obs_spectra}.

\begin{table}
\centering
\caption{\label{table:obs_spectra} Details of ESPaDOns spectral observations of Theta Eridani A.}
\begin{tabular}{cccc}
\hline \hline
\shortstack{date\\Time\\mean JD} & \shortstack{Exposure time\\Number of files} & \shortstack{$v_{\mathrm{z,Aa}}$\\($\text{ km}\text{ s}^{-1}$)} & \shortstack{$v_{\mathrm{z,Ab}}$\\($\text{ km}\text{ s}^{-1}$)} \\ [0.3cm]

\shortstack{2014-08-14\\15:11-15:25\\2456884.138} & \shortstack{40 s\\12} & $79.7\pm2.0$ & $-62.8\pm2.0$ \\ [0.3cm]

\shortstack{2014-11-04\\11:41-12:04\\2456965.996} & \shortstack{40 s\\16} & $63.6\pm2.0$ & $-46.9\pm2.0$ \\ [0.3cm]

\shortstack{2015-09-19\\10:45-11:28\\2457284.963} & \shortstack{45 s\\32} & $-65.7\pm2.0$ & $90.9\pm2.0$ \\ [0.3cm]

\shortstack{2015-10-29\\09:23-10:05\\2457324.906} & \shortstack{45 s\\32} & $6.3\pm2.0$ & $16.9\pm2.0$ \\ [0.3cm]

\hline
\end{tabular}
\end{table}

\subsection{FEROS}

We downloaded an archival FEROS \citep{Kaufer1999} spectra of $\theta$ Eri from the ESO Science Portal. It was taken in 2004-11-29 (MJD=53338.24) with an exposure time of 60 s, wavelength coverage 3600-9000{\AA} and spectral resolution $R=48,000$. The spectrum is that of an A-type star but it is not consistent with Theta Eridani A (the lines are much broader); therefore, we conclude that it is of Theta Eridani B. The spectrum was already corrected to the heliocentric frame. 

\subsection{TESS}

\begin{figure}[]
\centering
\includegraphics[width=0.5\textwidth]{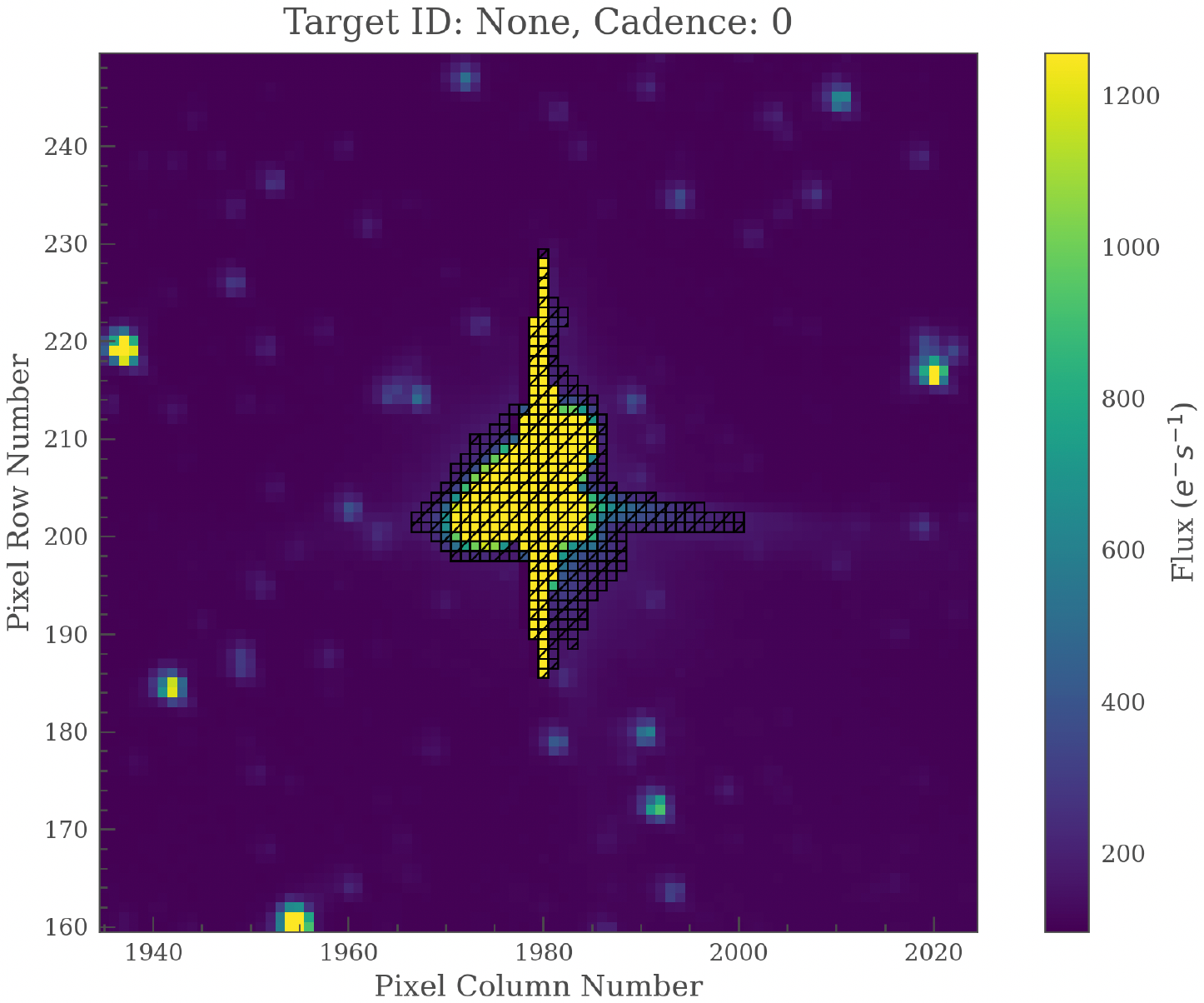}
\caption{\label{fig:tess_aperture} Example of a TESS Full Frame Image from Sector 31 centered on Theta Eridani. The aperture used to build the lightcurve is shown in black.}
\end{figure}

Theta Eridani was observed by TESS \citep{Ricker2015} during sectors 3 and 4 in 2018 with a cadence of 30 min and during sectors 30 and 31 in 2020 with a cadence of 10 min. Each sector comprises continuous observations over about 27 days. We downloaded 90x90 pixels TESS Full Frame Image (FFI) cutouts centered on $\theta$ Eri from the TESScut website \footnote{\href{https://mast.stsci.edu/tesscut}{https://mast.stsci.edu/tesscut}} \citep{Brasseur2019}. Figure \ref{fig:tess_aperture} shows an example FFI image from Sector 31 together with the aperture used (in black) for building the lightcurve. TESS is one of the few modern instruments that can deal with very bright stars by spilling their flux over a wider area. The downside is that due to the large size of each pixel (21") many other sources may also contribute to the flux. Fortunately, the background around $\theta$ Eri is relatively empty and free from bright stars. The combined Gaia DR3 \citep{Gaia2023} magnitude of all 331 sources within 11' of $\theta$ Eri is $R_p=9.32$, which is only 0.2\% of the flux from $\theta$ Eri ($R_p=2.72$), so that source contamination is negligible for our purposes (the $R_p$ filter of Gaia has a very similar bandpass to TESS). 

\begin{figure*}[]
\centering
\includegraphics[width=\textwidth]{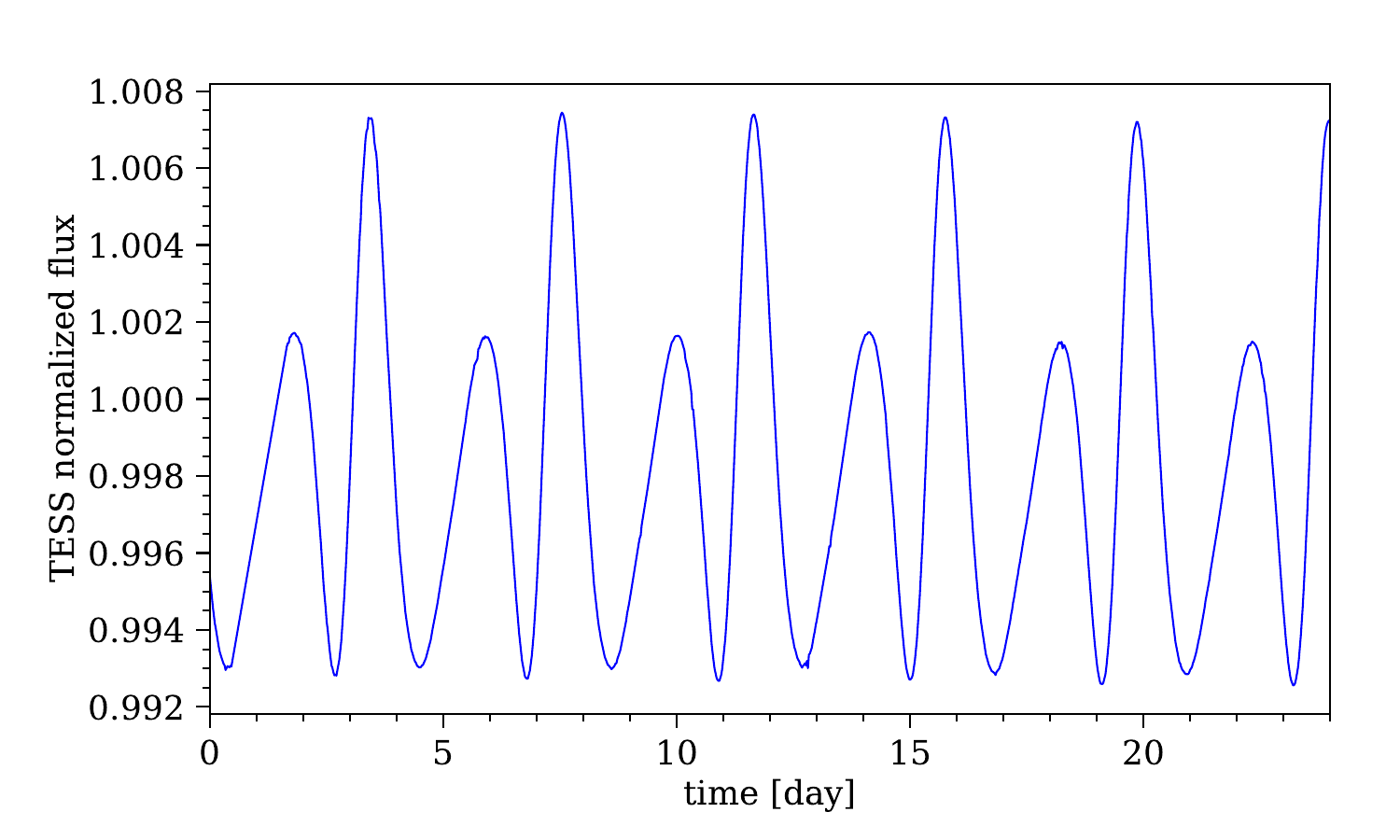}
\caption{\label{fig:tess_lightcurve_s3} TESS Sector 3 lightcurve of Theta Eridani.}
\end{figure*}

An example TESS lightcurve from Sector 3 is shown in Figure \ref{fig:tess_lightcurve_s3}. The lightcurves for the other sectors are virtually identical. 

\section{Results}
\label{sec:results}

\subsection{Separation vectors from VLTI/PIONIER}

We fit the PIONIER interferometric data with a binary model \citep[detailed in][]{Waisberg2023} consisting of the H band flux ratio ($\frac{f_{Ab}}{f_{Aa}}$) and the projected separation of Ab relative to Aa in the East and North directions $(\Delta \alpha_*, \Delta \delta)_{\mathrm{Ab,Aa}}$. The angular diameters of the stars were fixed to $\theta_{Aa} = 0.78 \text{ mas}$ and $\theta_{Ab} = 0.72 \text{ mas}$ based on the radii estimated below. They are too small compared to the interferometric resolution to be meaningfully constrained by the data. 

\begin{figure*}[]
\centering
\includegraphics[width=\textwidth]{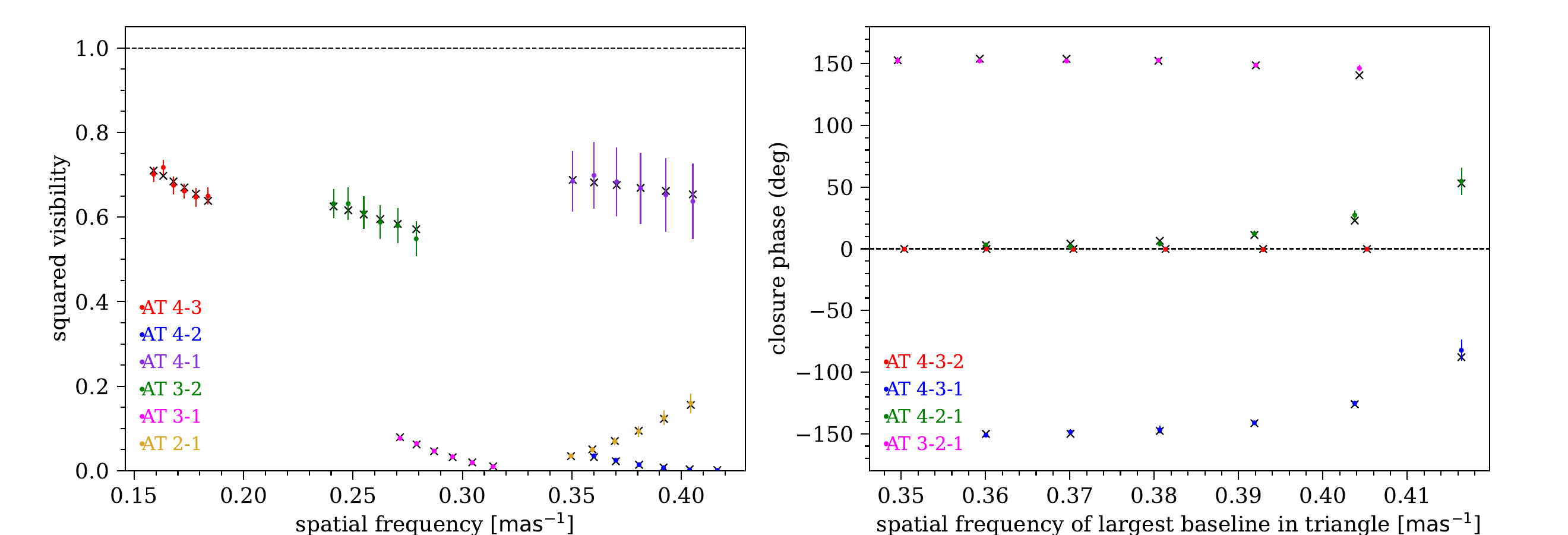}
\caption{\label{fig:pionier_fit_2017-08-27} VLTI/PIONIER data (colored) and best fit binary model (solid black) for $\theta$ Eri A for epoch 2017-08-27. The dashed lines show the expected values for a single unresolved star.}
\end{figure*}

Figure \ref{fig:pionier_fit_2017-08-27} shows the VLTI/PIONIER data and best fit binary model for epoch 2017-08-27. Corresponding figures for the other four epochs can be found in Appendix \ref{app:pionier_fits}. The binary model fit results for each epoch are reported in Table \ref{table:pionier_obs}. The formal astrometric errors from interferometric binary fits are usually underestimated due systematic errors from calibration and correlations between spectral channels. We adopted an astrometric error of $40 \mu$as in each direction based on the orbital fit residuals and our previous experience with interferometric data. 

\subsection{Orbital parameters}
\label{subsec:orbital_fit}

We performed a joint fit for the orbital parameters, individual dynamical masses and dynamical parallax for $\theta$ Eri Aa + Ab using the interferometric data and radial velocities (\ref{subsec:spectra}). In addition to the five VLTI/PIONIER epochs, we also used the VLTI/GRAVITY data point from \cite{Waisberg2025}, reproduced here for convenience: 

\begin{align}
\mathrm{MJD} &= 60549.297 \\
(\Delta \alpha_*, \Delta \delta)_{\mathrm{Ab,Aa}} &= (-0.24, 1.16) \text{ mas}
\end{align}

We adopted an astrometric error of $60 \mu$as for the VLTI/GRAVITY data due to its slightly lower spatial resolution compared to VLTI/PIONIER. 

The astrometric data constrain the seven Keplerian parameters, namely the angular semi-major axis $a$, the eccentricity $e$, the period $P$, the time of pericenter $T_p$, the argument of pericenter of the secondary $\omega$, the inclination $i$ and the longitude of the ascending node $\Omega$. The radial velocities constrain $e$, $\omega$, $P$, $T_p$ in addition to the systemic velocity $\gamma$ and the semi-amplitudes $K_{Aa}$ and $K_{Ab}$. The individual dynamical masses can then be obtained from 

\begin{align}
q = \frac{M_{Ab}}{M_{Aa}} = \frac{K_{Aa}}{K_{Ab}} \\
M_{Aa} = \frac{P}{2 \pi G} K_{Aa}^3 (1-e^2)^{3/2} \frac{(1+q)^2}{q^3}
\end{align}

The physical semi-major axis follows from Kepler's Third Law and together with the angular semi-major axis yields the distance (i.e. dynamical parallax).

\begin{figure*}[]
\centering
\includegraphics[width=\textwidth]{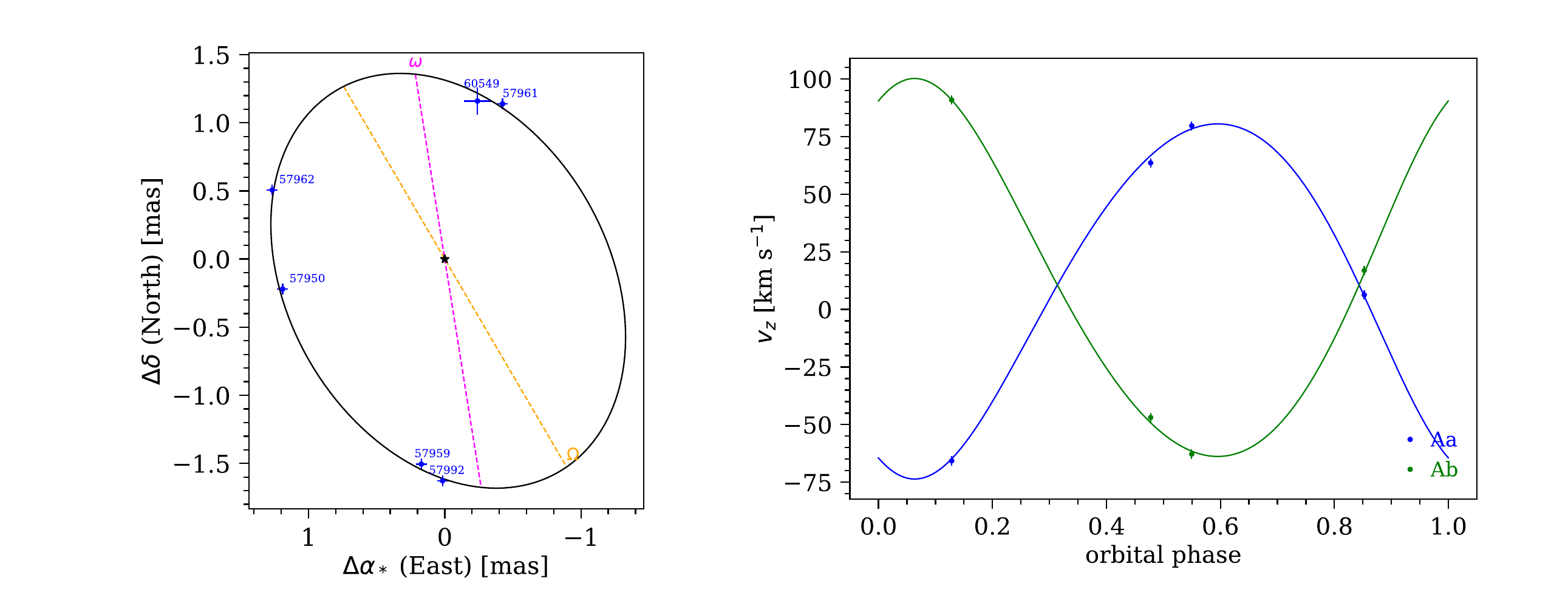}
\caption{\label{fig:orbital_fit} Joint astrometric and radial velocities fit for Theta Eridani Aa+Ab. \textbf{Left}: VLTI/PIONIER (labeled by MJD=57950 to 57992) and GRAVITY (MJD=60549) data (blue) and best fit orbit (black).  The star at (0,0) marks the position of the primary. The dashed magenta and orange lines show the line of apsides and the line of nodes respectively. \textbf{Right}: Radial velocity curves.}
\end{figure*}

In practice, we fix the period to the extremely precise value obtained from the TESS lightcurve (\ref{subsec:photometry}). We then find the best fit solution by running a series of nonlinear least squares fits over a finely sampled grid in $e$ and $T_p$. The uncertainties are estimated by generating 5000 resamples of the data according to the errors and finding the distribution of the best fit parameters. We report the median, 16\% and 84\% percentiles of the resulting distributions in Table \ref{table:orbital_fit}. The data and the best fit solution are plotted in Figure \ref{fig:orbital_fit}. The full parameter distributions can be found in Figure \ref{fig:corner}. There are no strong degeneracies except between $\omega$ and $T_p$. 

\begin{table}
\centering
\caption{\label{table:orbital_fit} Best fit parameters for Theta Eridani Aa+Ab.}
\begin{tabular}{ccc}
\hline \hline

& & source \\ 

\shortstack{$a$\\(mas)} & \shortstack{$1.62\pm0.03$} & astrometry \\ [0.3cm]

$e$ & $0.105\pm0.010$ & astrometry + RVs \\ [0.3cm]

\shortstack{$i$\\(deg)} & $43.2\pm2.3$ & astrometry \\ [0.3cm]

\shortstack{$\Omega$\\(deg)} & $30.7\pm1.4$ & astrometry \\ [0.3cm]

\shortstack{$\omega$\\(deg)} & $332.2\pm5.4$ & astrometry + RVs \\ [0.3cm]

\shortstack{$P$\\(days)} & $4.107704\pm0.000008$ & lightcurve \\ [0.3cm]

\shortstack{$T_p$\\(JD)} & $2457949.89\pm0.06$ & astrometry + RVs \\ [0.3cm]

\shortstack{$K_{Aa}$\\($\text{ km}\text{ s}^{-1}$)} & $77.0\pm1.4$ & RVs \\ [0.3cm]

\shortstack{$K_{Ab}$\\($\text{ km}\text{ s}^{-1}$)} & $81.9\pm1.4$ & RVs \\ [0.3cm]

\shortstack{$\gamma$\\($\text{ km}\text{ s}^{-1}$)} & $10.6\pm0.7$ & RVs \\ [0.3cm]

\shortstack{$d$\\(pc)} & $51.2\pm0.4$ & \shortstack{Gaia DR3\\$\theta$ Eri B} \\ [0.3cm]

\hline \\ [0.1cm]

\shortstack{$a$\\(au)} & $0.083\pm0.002$ & $a$(mas) and $d$ \\ [0.3cm]

$q=\frac{M_{Ab}}{M_{Aa}}$ & $0.94\pm0.02$ & $K_{Aa}$ and $K_{Ab}$ \\ [0.3cm]

\shortstack{$M_{Aa}$\\($M_{\odot}$)} & $2.33\pm0.13$ & $a$(au), $P$ and $q$ \\ [0.3cm]

\shortstack{$M_{Ab}$\\($M_{\odot}$)} & $2.19\pm0.13$ & $a$(au), $P$ and $q$ \\ [0.3cm]

\hline \\ [0.1cm]

\shortstack{$R_{Aa}$\\($R_{\odot}$)} & $4.3\pm0.1$ & lightcurve \\ [0.3cm]

\shortstack{$R_{Ab}$\\($R_{\odot}$)} & $3.95\pm0.1$ & lightcurve \\ [0.3cm]

\shortstack{$T_{Aa}$\\(K)} & $7600$ & space photometry + spectrum \\ [0.3cm]

\shortstack{$T_{Ab}$\\(K)} & $7800$ & space photometry + spectrum \\ [0.3cm]

\shortstack{$(v\sin i)_{Aa}$\\($\text{ km}\text{ s}^{-1}$)} & 70 & spectrum \\ [0.3cm]

\shortstack{$(v\sin i)_{Ab}$\\($\text{ km}\text{ s}^{-1}$)} & 60 & spectrum \\ [0.3cm]

\hline
\end{tabular}
\tablenotetext{0}{Notes:}
\tablenotetext{0}{The uncertainties correspond to the 16\% and 84\% percentiles of the parameter distributions.}
\end{table}

The combination of the astrometric and spectroscopic data yield dynamical masses $M_{Aa}=2.70_{-0.30}^{+0.36} M_{\odot}$ and $M_{Ab}=2.54_{-0.28}^{+0.34} M_{\odot}$ and a dynamical distance $d=53.7_{-2.6}^{+3.2} \text{ pc}$. The latter is consistent with albeit more uncertain than the Gaia DR3 distance of $\theta$ Eri B $d=51.2\pm0.4 \text{ pc}$, which is a single star as measured from interferometry \citep{Waisberg2025} and has a trustworthy Gaia parallax (RUWE=0.86). Therefore, in order to reduce the errors in the masses, for our official solution (reported in Table \ref{table:orbital_fit}) we adopted the Gaia DR3 distance and only use the radial velocity semi-amplitudes $K_{Aa}$ and $K_{Ab}$ to measure the mass ratio, with the total mass following from Kepler's Third Law. 

\subsection{Radii from ellipsoidal modulation}
\label{subsec:lightcurve}

The TESS lightcurve clearly shows that $\theta$ Eri is an ellipsoidal variable with a full variation in flux $2\frac{\Delta F}{F} \simeq 1.3\%$. It also confirms the small eccentricity found in the orbital solution, as there is a small asymmetry in the lightcurve that cannot be produced by a circular orbit. Based on the times of maxima we measured an orbital period $P=4.107704\pm0.000008 \text{ day}$. Moreover, an O-C analysis based on the total time span of 190 orbits covered by the TESS observations revealed no significant period change, with an upper limit $\dot{P} \lesssim 10^{-6}$. A Lomb-Scargle periodogram reveals no high frequency signals (such as rotation or pulsation) -- the lightcurve looks remarkably smooth. 

The amplitude of the ellipsoidal modulations of the primary scales as 

\begin{align}
\frac{\Delta F_{Aa}}{F_{Aa}} \propto \frac{M_{Ab}}{M_{Aa}} \left ( \frac{R_{Aa}}{a} \right )^3 \sin^2 i (1 + 2 k_{2,Aa}) 
\end{align}

\noindent \citep[e.g.][]{Mazeh2008}, where $k_2$ is the apsidal constant and an analogous expression applies to the secondary. Given that $k_2$ is negligible for both stars \citep[e.g. $k_2 \simeq 0.002$][]{Claret2023}, the ellipsoidal amplitude is a sensitive measurement of radii since the mass ratio, semi-major axis and orbital inclination are well-constrained by the astrometric plus spectroscopic orbital solution (\ref{subsec:orbital_fit}). In addition, given the radius of the primary Aa, the radius of the secondary Ab can be well approximated by 

\begin{align}
\frac{R_{Ab}}{R_{Aa}} \simeq f_{\mathrm{NIR}}^{1/2} \frac{T_{Aa}}{T_{Ab}} \simeq 0.88^{1/2} \times 0.974 \simeq 0.914 
\end{align}

\noindent where $f_{\mathrm{NIR}}$ is the near-infrared (H and K bands) flux ratio from the interferometry, $\frac{T_{Aa}}{T_{Ab}} \simeq \frac{7600}{7800}$ (\ref{subsec:photometry}) and the NIR is at the Rayleigh-Jeans tail of the stars' spectral energy density. Therefore, the amplitude of the ellipsoidal modulation can be used to measure both stellar radii. 

We modeled the binary lightcurve using the software PHOEBE \citep{Prsa2005,Prsa2016}. The orbital parameters and masses were fixed by the orbital solution (\ref{subsec:orbital_fit}). The temperatures were set to $T_{Aa}=7600 \text{ K}$, $T_{Ab}=7800 \text{ K}$ (\ref{subsec:photometry}). The stellar spin axes were set to be aligned with the orbital axis and the synchronization parameters were set to $F_{Aa}=F_{Ab}=2$ (\ref{subsec:spectra}). The limb darkening coefficients were set to the default values in PHOEBE, while the gravity darkening coefficients were set to 1 and the irradiation efficiency to 1 as appropriate for stars with a radiative envelope. These parameters have subtle effects on the lightcurve but the ellipsoidal amplitude is dominantly dependent on the stellar radii (for fixed $a$ and $i$). It was also important to set the ``third light'' contamination fraction due to component $\theta$ Eri B to 33\% based on $\Delta R_p = 1.19$ between B and A from Gaia DR3 (the $R_p$ filter is very similar to the TESS filter). 

\begin{figure}[]
\includegraphics[width=\columnwidth]{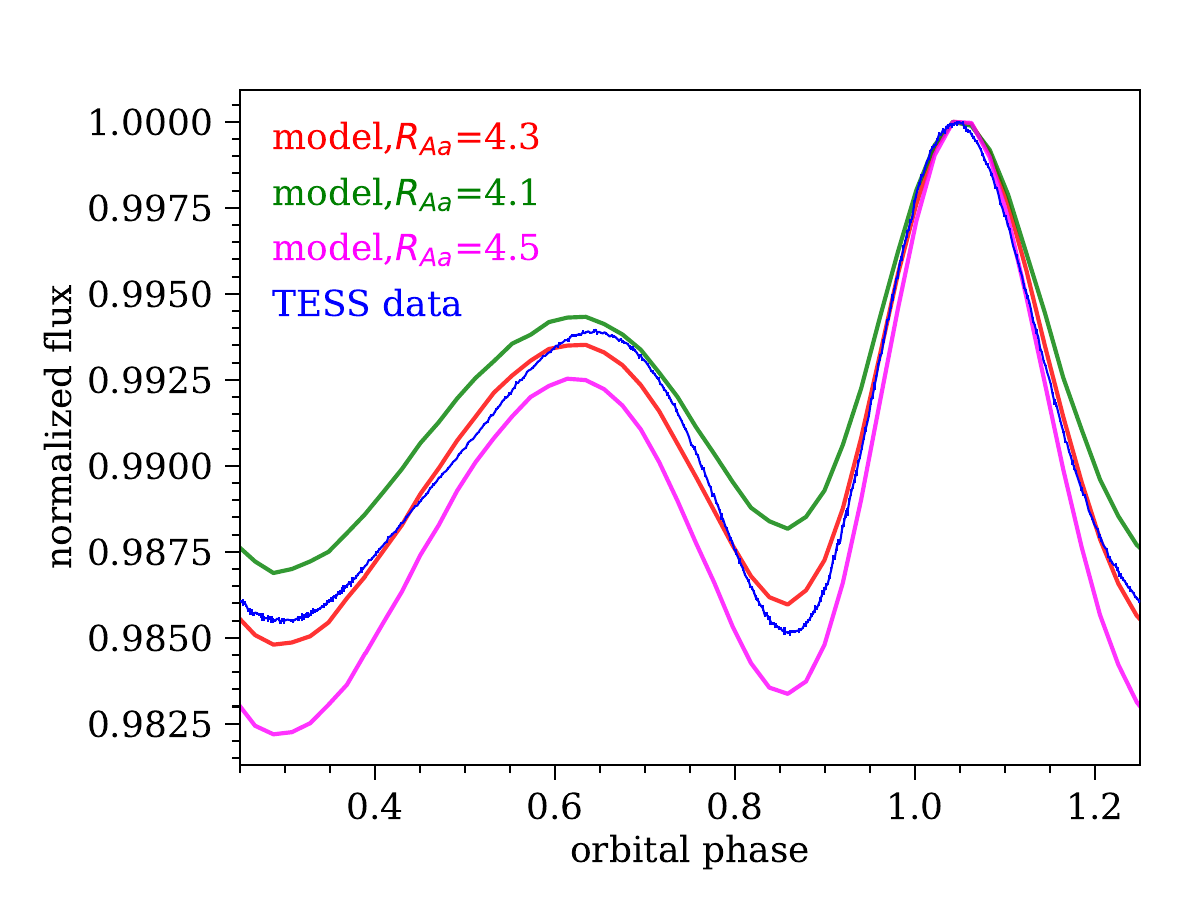}
\caption{Phase-averaged TESS lightcurve of Theta Eridani (blue) and corresponding PHOEBE lightcurve models for different values for the radius of the primary star Aa. Both data and models are normalized so that the peak flux is 1.0.}
\label{fig:phoebe_model}
\end{figure}

Figure \ref{fig:phoebe_model} shows the phase-averaged TESS lightcurve of Theta Eridani (blue), where phase 1.0 corresponds to periastron passage, and the corresponding PHOEBE models for $R_{Aa}=$4.1, 4.3 and 4.5 $R_{\odot}$. Both data and models are normalized so that the peak is at a flux of 1.0. Allowing $a$ and $i$ to vary within $1\sigma$ of the orbital solution, we find $R_{Aa}=4.3\pm0.1 R_{\odot}$. The corresponding secondary radius is $R_{Ab}=4.0\pm0.1 R_{\odot}$. 

We note that the orbital and stellar parameters can potentially be refined by a joint fit of the astrometry, radial velocities and lightcurve, but this is beyond the scope of this paper. 

\subsection{Photometric temperatures}
\label{subsec:photometry}

We estimated the effective temperatures of Aa and Ab by comparing space photometry of $\theta$ Eri A against PHOENIX atmosphere models from the Göttingen Spectral Library \footnote{\url{https://phoenix.astro.physik.uni-goettingen.de/?page_id=15}} \citep{Husser2013}. The models have steps $\Delta \log g = 0.5$ and $\Delta T = 200 \text{ K}$ and solar metalicity. From the masses and radii measured above we have $\log g_{Aa}=3.54$ and $\log g_{Ab}=3.57$, so we used the models with $\log g=3.5$. The radii were fixed to $R_{Aa}=4.3 R_{\odot}$ and $R_{Ab}=4.0 R_{\odot}$, the distance to $d=51.2 \text{ pc}$ and we fit for the two temperatures $T_{Aa}$ and $T_{Ab}$.  

We used space photometry from Tycho2 \citep[$V_T=3.195\pm0.009$, $B_T=3.370\pm0.014$;][]{Hog2000} and WISE \citep[$W_1=2.846\pm0.423$, $W_2=1.996\pm0.268$;][]{Cutri2012}, which report separate measurements for $\theta$ Eri A and B. The model fluxes over each filter were calculated using the public \texttt{python} module \texttt{pyphot} \citep{Fouesneau20215}. 2MASS reports only a joint (and very uncertain) measurement for A+B and is therefore not useful. We also attempted to use Gaia photometry (which dooes report separate measurements for A and B) but we found that the Gaia errorbars are severely underestimated and the residuals were very high for the $g$ ($\sim10\%$) and $b_p$ ($\sim25\%$) filters. A comparison between Gaia DR2 and DR3 shows discrepant $g$ magnitudes by 0.1 and a completely wrong $b_p$ magnitude in DR2. This is likely related to the current Gaia data release not being optimized for very bright stars. 

\begin{figure}[]
\includegraphics[width=\columnwidth]{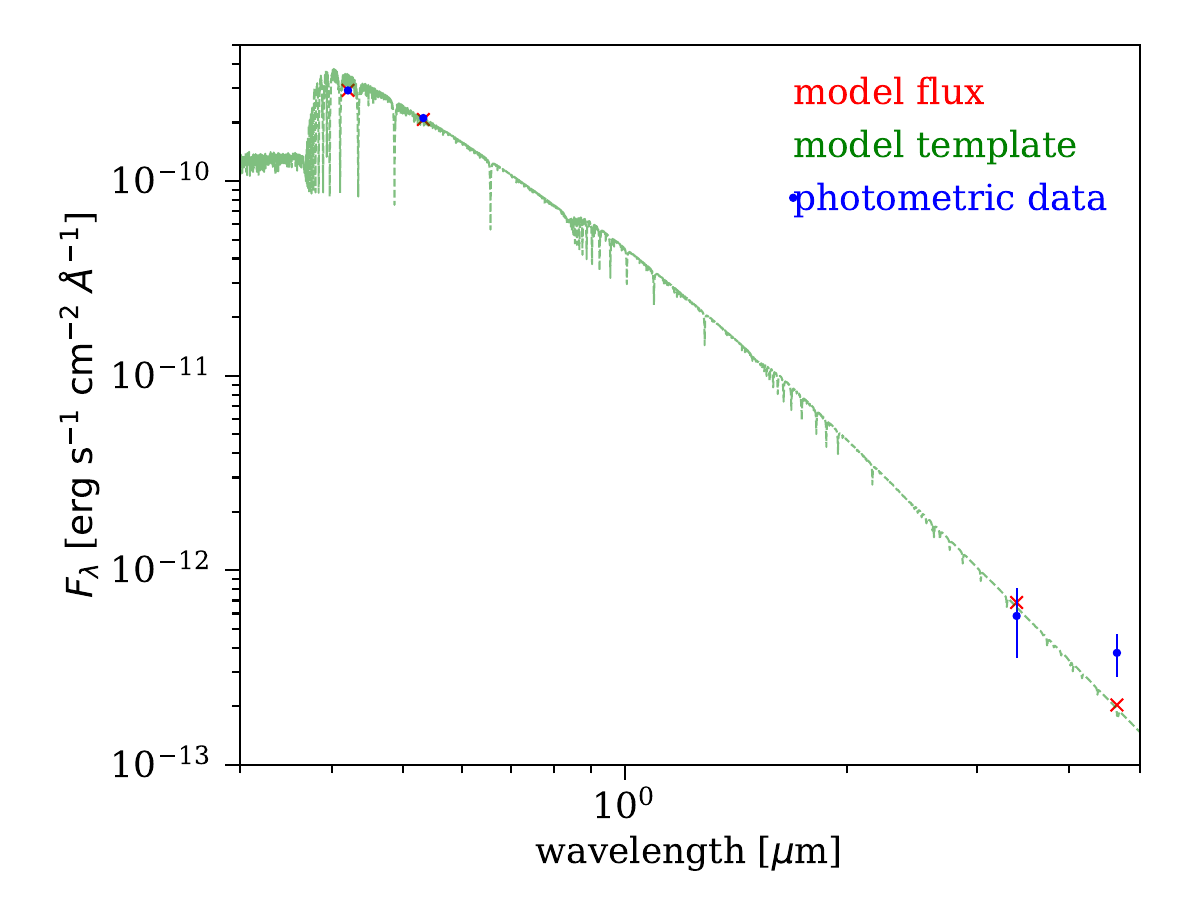}
\caption{Photometric data of Theta Eridani A (blue; from left to right: $B_T$, $V_T$, $W_1$, $W_2$) and the best fit model (green) and corresponding photometry (red) consisting of the sum of PHOENIX models for the primary ($T_{Aa}=7600 \text{ K}$, $\log g =3.5$, $R_{Aa}=4.3 R_{\odot}$) and for the secondary ($T_{Ab}=7800 \text{ K}$, $\log g =3.5$, $R_{Aa}=4.0 R_{\odot}$).}
\label{fig:photometry}
\end{figure}

We found that the two acceptable solutions are $(T_{Aa},T_{Ab}) \simeq (7600,7800) \text{ K}$ or $(T_{Aa},T_{Ab}) \simeq (7800,7600) \text{ K}$. As can be seen in \ref{subsec:spectra}, the spectra of the two stars are almost identical but the primary has a slightly stronger Ca II 3933{\AA} absorption line (Figure \ref{fig:spectrum_1}), which implies that it is just slightly cooler. Therefore, we adopt the former solution for the temperatures. Figure \ref{fig:photometry} shows the photometric data (blue) together with the best fit model (green) and corresponding photometry (red) consisting of the sum of PHOENIX models for the primary ($T_{Aa}=7600 \text{ K}$, $\log g =3.5$, $R_{Aa}=4.3 R_{\odot}$) and for the secondary ($T_{Ab}=7800 \text{ K}$, $\log g =3.5$, $R_{Aa}=4.0 R_{\odot}$). We did not include the WISE bands W3 and W4 in the fit because the models do not extend to their wavelengths, but we note that an extrapolation from the model is consistent with the measurements and therefore there is no evidence for any infrared excess despite the somewhat large (1.8$\sigma$) positive residual in W2. 
 
\subsection{Spectral analysis}
\label{subsec:spectra}

\subsubsection{Theta Eridani A}

\begin{figure*}[]
\includegraphics[width=2\columnwidth]{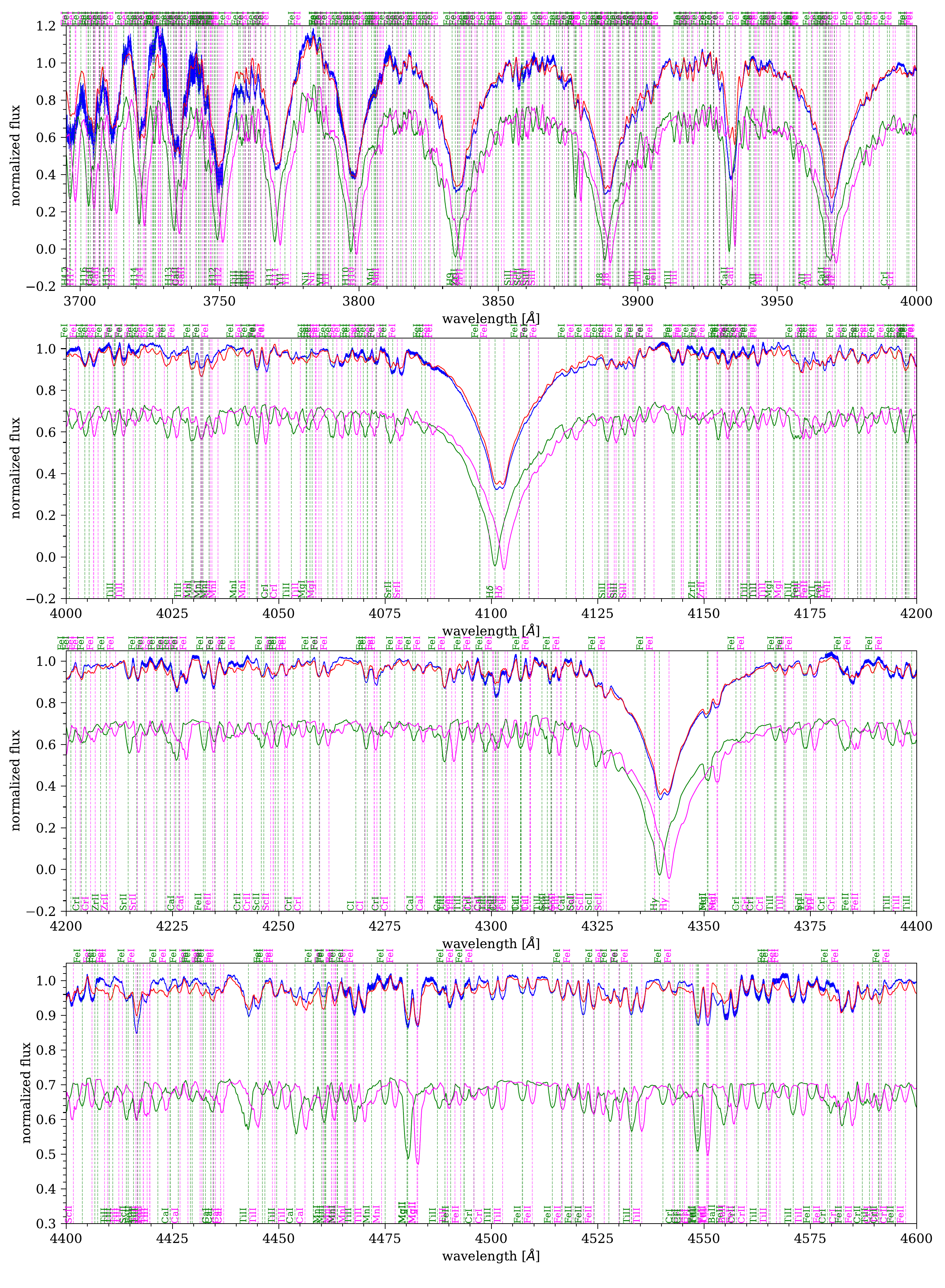}
\caption{Espadons normalized spectrum of Theta Eridani A (blue) for epoch 2015-09-19 (maximum velocity separation) together with normalized PHOENIX model spectra for the primary Aa (green), secondary Ab (magenta) and their sum (red). The SB2 nature of the star is clearly apparent in the stronger absorption lines. The magenta and green lines are shifted vertically by -0.3 for clarity.}
\label{fig:spectrum_1}
\end{figure*}

Figures \ref{fig:spectrum_1}, \ref{fig:spectrum_2}, \ref{fig:spectrum_3} and \ref{fig:spectrum_4} show the normalized Espadons spectrum (blue) of $\theta$ Eri A for epoch 2015-09-19, in which the two components have the largest velocity separation. The SB2 nature of the star is clearly apparent in the stronger absorption lines. Normalized PHOENIX model spectra for the primary Aa (green) and secondary Ab (magenta) with parameters obtained in the previous sections as well as their sum (red) are also shown. A detailed inspection of the spectrum reveals that the components have virtually identical spectra, but the primary is just slightly cooler based on its slightly stronger Ca II 3933{\AA} line. The vertical lines mark the identified spectral lines for the primary and secondary in green and magenta, respectively (Fe I lines, the most numerous by far, are marked on top). Table \ref{table:line_list} lists all the identified lines. 

We measured the projected rotational velocities of both components by rotationally broadening the model spectra and matching them against the observed spectrum. We found $(v\sin i)_{Aa} \simeq 70 \text{ km}\text{ s}^{-1}$ and $(v\sin i)_{Ab} \simeq 60 \text{ km}\text{ s}^{-1}$. If the stellar spin axes are aligned with the orbital axis, the rotation periods of the stars are then $P_{\mathrm{rot,Aa}} \simeq P_{\mathrm{rot,Ab}} \simeq 2 \text{ days}$. The radial velocities of both components were them measured by fitting the template spectra against each observed spectrum across a 2d grid in the velocity shifts. We estimated an error of $2 \text{ km}\text{ s}^{-1}$ based on the scatter of the results across different spectral regions rich in absorption lines. The best fit radial velocities are reported in Table \ref{table:obs_spectra}. 

Because the two stars have virtually identical spectra, the cleanest spectrum to assess possible chemical anomalies is that of epoch 2015-10-29, in which the velocity separation is minimal and the spectrum looks like that of a single star. We plot such spectrum in Figures \ref{fig:spectrum_single1}, \ref{fig:spectrum_single2}, \ref{fig:spectrum_single3} and \ref{fig:spectrum_single4}. Through visual inspection we find that, within the limitations of the atmosphere model \footnote{Namely, a fixed microturbulent velocity of $0.8 \text{ km}\text{ s}^{-1}$, the assumption of local thermodynamic equilibrium and of a single temperature, when in reality the temperature of the star varies with latitude due to gravity darkening caused by its non negligible rotation.}, the vast majority of elements have surface abundances quite close to solar. The only exception are s-process elements, whose surface abundance appears enhanced (see e.g.
Sr II 4077.7{\AA}, Sr II 10327.3{\AA}, Y II 4883.7{\AA}, YII 4900.1{\AA}, Ba II 4934.1{\AA}, Ba II 5853.7{\AA}, BaII 6141.7{\AA}). While such enhancement is typical of chemically peculiar Am stars, we dot not find the flagship marks of such stars such as under-abundant Ca and Sc and over-abundant iron-peak elements. This is consistent with the rotational velocities being large enough ($v_{Aa} \simeq 100 \text{ km}\text{ s}^{-1}$ and $v_{Ab} \simeq 90 \text{ km}\text{ s}^{-1}$ in case of spin-orbit alignment) to not allow the development of strong chemical peculiarities from radiative diffusion \citep{Abt1995}.
Because the spectrum of wide companion Theta Eri B is also found to be enhanced in s-process elements (\ref{subsec:spectra_B}), we conclude that the most likely scenario is that the Theta Eridani system was formed from gas that was somewhat enriched in s-process elements from a previous generation of AGB star(s). 

\subsubsection{Theta Eridani B}
\label{subsec:spectra_B}

Figures \ref{fig:spectrum_B1}, \ref{fig:spectrum_B2}, \ref{fig:spectrum_B3} and \ref{fig:spectrum_B4} show the normalized FEROS spectrum of Theta Eridani B (blue) together with a PHOENIX model atmosphere with $\log g=4.0$ and $T=8200 \text{ K}$, according to the isochrone photometric parameters in \cite{Waisberg2025}. The same list of spectral lines as in $\theta$ Eri A are marked. 

Compared to $\theta$ Eri Aa and Ab, $\theta$ Eri B has a larger projected rotational velocity $(v \sin i)_B \simeq 100 \text{ km} \text{ s}^{-1}$ that makes its metal lines shallower and broader. Interestingly, the spectrum also shows weak and much narrower absorption lines superposed onto the broad photospheric lines. Given that Theta Eri B has no detected interferometric companion \citep{Waisberg2025} and the narrow absorption lines are at the star's systemic velocity $\gamma_B \simeq 8\pm1 \text{ km}\text{ s}^{-1}$, they likely arise from an extended equatorial outflow, which makes $\theta$ Eri B a mild shell star. Shell spectra are very common among fast rotating A-type stars (e.g. \textit{Vega}), suggesting that the true rotational velocity of $\theta$ Eri B is probably significantly higher than $(v \sin i)_B$. Due to the large temperature variation from the pole to the equator caused by gravity darkening, abundances assessment is considerably more difficult and requires more detailed multi-temperature modeling. However, as is the case for $\theta$ Eri A, several lines from s-process elements (Sr II, Y II, Ba II) are significantly enhanced. Such peculiarity cannot be explained by the Am phenomenology in the case of $\theta$ Eri B and support the case for an intrinsic s-process enhancement for the $\theta$ Eri system.

\section{Discussion}
\label{sec:discussion}

\begin{comment}
\subsection{One of the brightest ellipsoidal variables}

There are currently 267 confirmed or candidate rotating ellipsoidal variables (ELL) in the General Catalogue of Variable Stars \citep{Samus2017} and only one of them is brighter than $\theta$ Eridani A (namely, \textit{Spica})\footnote{Gamma Lupi (V=2.7) is listed as a candidate ELL but was found not to have ellipsoidal variations in \cite{Jerzykiewicz2021}.}. Considering the 6,055 Algol-type (EA) and 1,834 Beta Lyrae-type (EB) eclipsing systems, there are only a further five systems with ellipsoidal variations that are brighter than $\theta$ Eri A (namely, \textit{Algol}, \textit{Mintaka}, $\delta$ Cap, $\pi$ Sco and $\mu^1$ Sco). 
\end{comment}

Was Theta Eridani much brighter in the past? The current luminosity of the Theta Eridani system is $L = L_{Aa} + L_{Ab} + L_{B} \simeq (56 + 52 + 44) L_{\odot} = 152 L_{\odot}$ and the SED peaks relatively close to the V band (Figure \ref{fig:photometry}). Therefore, if the system was about ten times brighter in the V band in the past, it must have had a luminosity of at least about $L_{\mathrm{outburst}} \sim 1500 L_{\odot}$ (potentially higher if the emission was hotter). In order to sustain this luminosity for at least 1000 years, the total energy required is $E \sim 2 \times 10^{47} \text{ erg}$. Incidentally, this is comparable (and smaller) than the orbital energy of the close binary Aa+Ab

\begin{align}
E_{\mathrm{orb}} = \frac{G M_{Aa} M_{Ab}}{2 a} \simeq 5.4 \times 10^{47} \text{ erg}
\end{align}

\noindent suggesting that extraction of its orbital energy is a possible source for the brightening. 

In order to explore a possible mechanism for the transient, we note a further two tantalizing facts about the binary. 

%The current absolute visual magnitude of Theta Eridani is $M_V \simeq -0.7$ so in order to explain its brightening by $\Delta V \sim 2.7$ the total visual magnitude of the system should be $M_V \simeq -3.4$ and that of the accretion disk alone $M_V \simeq -3.3$. 

The first is that the primary Aa is close to filling its Roche lobe. This is already clear from the fact that the ellipsoidal variations are rather large (Figure \ref{fig:tess_lightcurve_s3}). Quantitatively, the standard \cite{Eggleton1983} formula for the volume-equivalent radius of the Roche lobe for a mass ratio $q=0.94$ gives $\frac{R_{L,Aa}}{a} = 0.384$, where $a$ is the semi-major axis in a circular binary. Although the Roche formalism does not strictly apply to an eccentric binary, at pericenter we have $R_{L,Aa} \simeq 0.384 \times (1-0.105) \times 0.083 \text{ au} = 6.1 R_{\odot}$. However, the \cite{Eggleton1983} formula also assumes synchronization between the stellar rotation and the orbital period. For a super-synchronous star the Roche lobe shrinks due to the extra centrifugal potential. In the case where the spin axes are aligned with the orbital axes, the stellar rotational frequencies are about two times the orbital frequency (\ref{subsec:spectra}) and PHOEBE reports that the effective Roche lobe radii at periastron are actually $R_{L,Aa} = 5.6 R_{\odot}$ and $R_{L,Ab} = 5.4 R_{\odot}$; therefore, the radii $R_{Aa} \simeq 4.3 R_{\odot}$ and $R_{Ab} \simeq 4.0 R_{\odot}$ are about 80\% and 74\% of their Roche lobe radii at periastron, respectively.

\begin{figure*}[]
\includegraphics[width=2\columnwidth]{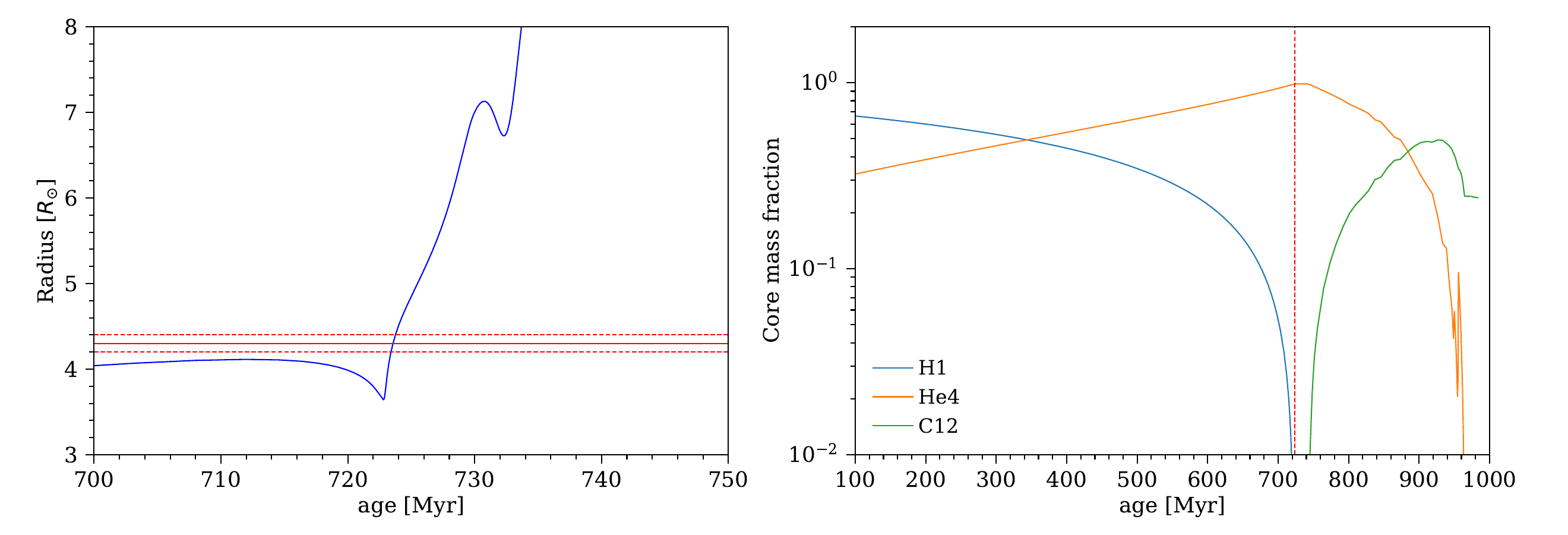}
\caption{Evolution of a $2.30 M_{\odot}$ star from a \texttt{MIST} EEP file. Theta Eridani Aa, with a radius $R_{Aa}=4.3\pm0.1 R_{\odot}$, is at the very beginning of the subgiant phase just after hydrogen core exhaustion.}
\label{fig:evolution}
\end{figure*}

The second intriguing fact is that the primary appears to be at a very special stage in its evolution, namely at the very beginning of the subgiant phase i.e. when the star has just finished core hydrogen burning. Figure \ref{fig:evolution} shows the evolution of the stellar radius (left) and of the core mass fractions of hydrogen, helium and carbon (right) for a $2.30 M_{\odot}$ star from a MESA Isochrones and Stellar Tracks \citep[\texttt{MIST};][]{Choi16,Dotter16,Paxton11,Paxton13,Paxton15} Equivalent Evolutionary Points (EEP) file. In the left plot, the red lines mark the current radius of the primary $R_{Aa} = 4.3\pm0.1 R_{\odot}$. In the right plot, the red line marks the age of 725 Myr corresponding to a $4.3 R_{\odot}$ radius. Hydrogen is virtually completely exhausted in the core with a mass fraction $2\times10^{-9}$. 

A detailed model that explains the transient and its properties is beyond the scope of this paper. Below we outline a possible mechanism in which the transient is powered by orbital energy extraction.

First, we note that for a peak temperature $T_{\mathrm{outburst}} \sim 8000 \text{ K}$ close to the V band, the radius of the emitting region during the outburst

\begin{align}
R_{\mathrm{outburst}} \sim \left ( \frac{L_{\mathrm{outburst}}}{4 \pi \sigma T_{\mathrm{outburst}}^4} \right )^{1/2} \simeq 20 R_{\odot}
\end{align}

\noindent where $\sigma$ is the Stefan-Boltzmann constant. A higher temperature can reduce the required outburst radius but too high a temperature would lead to an unphysically high bolometric luminosity. For example, to have $R_{\mathrm{outburst}} \simeq 6 R_{\odot}$ (the approximate size of the Roche lobe radius) and an increase in the V band flux by a factor of 8 would require a blackbody temperature $T_{\mathrm{outburst}} \simeq 20000 \text{ K}$, resulting in a bolometric luminosity $L_{\mathrm{outburst}} \simeq 5000 L_{\odot}$, which would require a total energy over 1000 yrs that is larger than the total orbital energy of the system. Therefore, it is likely that the outburst was produced within an extended envelope that engulfed the binary system. 

How could such a configuration develop? We suggest that prior to the transient the orbit had a larger semi-major axis and eccentricity. At some point, the primary star filled its ``Roche lobe'' at periastron, triggering mass transfer. The latest stages of the expansion of the primary could have been caused either by stellar evolution, tidal dissipation or both. For eccentric binaries there is no rotating frame in which the equipotential surfaces are time independent. As a result, the mass transfer process is much more complex than in the circular case and our understanding of the so-called ``eccentric Roche-lobe overflow'' is still quite limited. In any case, attempts to model the process using smoothed particle hydrodynamics simulations find that the formation of a common envelope is a possible outcome for significantly eccentric binaries \citep[e.g.][]{Regos2005,Lajoie2011}. 

In order for the process to last over $n \sim \frac{1000 \text{ yrs}}{4 \text{ days}} \sim 10^{5}$ orbits, the mass in the circumbinary envelope must be much smaller than that of the system. In this case, it is likely that a non-negligible fraction of the circumbinary mass is ejected every orbit due to the instability of a test particle in this equal mass eccentric binary. The total mass ejection is limited by the available energy to be less than about $M_{\rm ej}\sim 0.05 M_{\odot}$. This implies that the circumbinary mass at any given time is of order $M_{\rm env}\sim \frac{M_{\rm ej}}{n}\sim 5\times 10^{-7} M_{\odot}$. The optical depth of such an envelope is large

\begin{align}
\tau\sim\kappa \frac{ M_{\rm env}}{4\pi R_{\rm env}^2}\sim 80 \left(\frac{\kappa}{\rm cm^2/g}\right) \left(\frac{M_{\rm env}}{10^{-6} M_{\odot}}\right)\left(\frac{R_{\rm env}}{20 R_{\odot}}\right)
\end{align}

\noindent where $\kappa$ is the opacity. This is consistent with the requirement for emission from a large area. 

Because the envelope mass is very small compared to the binary, it is unlikely to be the source of orbital energy extraction through dynamical friction. Therefore, this is a different situation than a traditional ``common envelope'' that hardens a binary within a handful of orbits. In this case, we suggest that the orbital energy extraction is actually achieved through dynamical tidal dissipation arising from the order unit perturbations within each orbit. The envelope serves ``merely'' to reprocess the energy. It should also be noted that a fraction of the energy that powers the transient may also come from accretion of some of the matter lost by the primary onto the secondary star. 

Given that only a small mass is lost from the system, the angular momentum (which is dominated by the orbital angular momentum; the spin angular momentum is only about 7\% of the orbital angular momentum in the current binary) is approximately conserved. This implies that as the eccentricity goes down due to tidal dissipation, the pericenter goes up by a few tens of percents (since the approximtely conserved orbital angular momentum $\propto \sqrt{r_p(1+e)}$, where $r_p$ is the pericenter distance). This can provide a natural way for the mass transfer to stop. In this case a non zero residual eccentricity is expected. 

A further interesting clue comes from the fact that the stars have similar values for the observed $\frac{v \sin i}{R}$. An intriguing possibility is that the stellar spins were pseudo-synchronized with the previous more eccentric orbit and maintained their stellar angular momenta. In this case they would both have the same $i$ and period. If true, the current spin provides more information about the orbit before the transient started. 

The current rotation periods are $P_{\mathrm{rot,Aa}} \simeq P_{\mathrm{rot,Ab}} \simeq 2 \text{ days}$ (assuming the stellar spin axes are aligned with the orbital axis). The pseudo-synchronization period for an eccentric orbit is given by 

\begin{align}
P_{\mathrm{p-syn}} = \frac{f_5(e^2) (1-e^2)^{3/2}}{f_2(e^2)} P_{\mathrm{orb}}  \\
f_2(e^2) = 1 + \frac{15}{2} e^2 + \frac{45}{8} e^4 + \frac{5}{16} e^6 \\
f_5(e^2) = 1 + 3 e^2 + \frac{3}{8} e^4 
\end{align}

\noindent \citep{Hut1981}\footnote{We note that although the pseudo-synchronization period in \cite{Hut1981} comes from the theory of equilibrium tides, detailed calculations for dynamical tides (more relevant for the radiative envelopes in the case of Theta Eridani) find very similar results, e.g. expression (39) in \cite{Su2022} differs from \cite{Hut1981} only by a constant factor of 0.86.}. This can be simplified using Kepler's Third Law and conservation of total angular momentum to 

\begin{align}
P_{\mathrm{p-syn}} = \frac{f_5(e_i^2)}{f_2(e_i^2)} \frac{2\pi}{(G M_A)^{1/2}} (a_f (1-e_f^2))^{3/2}
\end{align}

\noindent where $a_f = 0.083 \text{ au}$ and $e_f = 0.105$ refer to the current (final) orbit. We can thus solve for the initial eccentricity $e_i$ such that $P_{\mathrm{p-syn}} = 2 \text{ days}$. We find $e_i \simeq 0.58$. 

From conservation of orbital angular momentum, this implies an initial semi-major axis $a_f \simeq 0.124 \text{ au}$ (and the corresponding orbital period $P_f \simeq 8 \text{ days}$). The extracted orbital energy is hence

\begin{align}
\Delta E_{\mathrm{orb}} = \frac{G M_{Aa}M_{Ab}}{2} \left ( \frac{1}{a_f} - \frac{1}{a_i} \right ) \approx 1.8 \times 10^{47} \text{ erg}
\end{align}

\noindent which is just what is needed to power the transient for 1000 years. The pseudo-Roche lobe size of the primary at periastron for such an orbit $r_L \simeq 0.384 a_i (1-e_i) = 4.3 R_{\odot}$ is equal to its current radius, implying that it would indeed be subject to eccentric Roche lobe overflow.

\subsection{How rare was the Theta Eridani outburst?}

The chance of observing the brightening of Theta Eridani within the binary's lifetime is $p \sim \frac{10^3 \text{ yrs}}{1 \text{ Gyr}} \sim 10^{-6}$. Therefore, it is hard to escape the fact that this was a rather remarkable event to happen within one of the $\sim 1000$ naked eye stars, especially considering that only a small minority of them are close eccentric binary systems that would be susceptible to our proposed mechanism. Incidentally, it is interesting to note that one of the stars close to Theta Eridani in Figure \ref{fig:historical_al-sufi}, namely HIP 42726 = HY Vel with mS=3 and modern V=4.83, is also a close, eccentric binary system \citep[$P=8.4\text{ days}$; $e=0.24$][]{DeCat2000}. 

Although modern photometric surveys have had an observational coverage of a few years to a few decades rather than two millennia like Theta Eridani, they are sensitive to objects up to 14 magnitudes fainter than the naked eye limit. Therefore, there could be unexplored or unexplained Theta Eridani-like transients that either brightened or faded in current data. In addition, if the typical outburst duration is indeed of the order of a millennium, there should be even more systems that are currently in outburst. However, correctly identifying them may be a challenge since it requires a detailed characterization of the system in order to establish a mismatch between its luminosity and that expected for the stars' masses. 

\section{Conclusion}
\label{sec:conclusion}

Theta Eridani was listed by both Ptolemy in 137 AD and by al-Sufi in 964 AD among the thirteen brightest stars in their (visible) night sky, in addition to being reported by Hipparchus around 129 BC as a particularly bright star. This is in stark contrast with its modern and relatively humble V=2.9 brightness. The visual magnitude discrepancy (after calibration of the ancient magnitudes to the modern scale) is $\Delta V = 2.7$ for Ptolemy's \textit{Almagest} (the highest among all stars) and $\Delta V = 2.4$ for al-Sufi's \textit{The Book of Fixed Stars} (the second highest among all stars). A visual magnitude V=3 consistent with the modern value was first reported from observations taken during the \textit{Eerste Schipvaart} expedition in 1598. The discrepancy with ancient observations has been a subject of controversy for over a century \citep{Anderson1893}.

We find no compelling reason to doubt the accuracy of the ancient records. In particular, explanations such as a confusion with Alpha Eridani or transcription/translation errors can be excluded. Although it is ultimately impossible to completely exclude the possibility of errors due to an over-correction of atmospheric extinction, we find that this hypothesis is not supported by the overall catalog data and would have to be a rather uncharacteristic error made by three independent ancient observers. 

Theta Eridani is actually a triple system. By combining interferometric, spectroscopic and photometric observations, we show that the visual primary is a close ($a=0.083 \text{ au}$, $P=4.10 \text{ days}$), mildly eccentric ($e=0.105$) binary. In particular, there are two rather tantalizing facts in light of the possible historical brightening: the stellar radii are $\sim 80\%$ of the Roche lobe radii and the primary has apparently just finished core hydrogen burning, with a radius $R_{Aa} \simeq 4.3 R_{\odot}$ that has doubled compared to its zero-age main sequence radius.

Motivated by the fact that the minimum energy required to power the putative transient is smaller but comparable to the orbital energy of Theta Eridani Aa+Ab, we propose that the brightening event was triggered when the primary star filled its ``Roche lobe'' in a previous more eccentric orbit. The orbital energy extracted through dynamical tidal dissipation was reprocessed by a light but optically thick outflowing circumbinary envelope ($M_{\mathrm{env}} \sim 5 \times 10^{-7} M_{\odot}$, $R_{\mathrm{env}} \sim 20 R_{\odot}$) with a total mass loss of the order $0.05 M_{\odot}$ over 1000 yrs. 

This model naturally accounts for the residual eccentricity of the current binary. The fact that the stars currently have a super-synchronous rotation and that their rotation periods are apparently the same could be explained if they were pseudo-synchronized to the previous orbit with $e \sim 0.6$ and $a \sim 0.124 \text{ au}$. An important remaining question is what sets the timescale of the transient or, equivalently, its luminosity. This is a potentially complicated question as it ties to the strength and evolution of dynamical tides during eccentric Roche lobe overflow. The discovery and characterization of more binary systems undergoing such process in modern photometric surveys holds the potential to better understand what may be a ubiquitous, short-lived but determinant phase in the evolution of close binaries. 

\section*{Acknowledgments}

This research has made use of the Jean-Marie Mariotti Center \texttt{SearchCal} service \footnote{Available at http://www.jmmc.fr/searchcal} co-developped by LAGRANGE and IPAG, CDS Astronomical Databases SIMBAD and VIZIER \footnote{Available at http://cdsweb.u-strasbg.fr/} and NASA's Astrophysics Data System Bibliographic Services, NumPy \citep{van2011numpy} and matplotlib, a Python library for publication quality graphics \citep{Hunter2007}. 

\section*{Data availability}
The data underlying this article is publicly available from the respective archives. Reduced data can be provided by the authors on request. 

\bibliographystyle{aasjournal}
\bibliography{main}{}

\appendix

\section{A. VLTI/PIONIER data and binary fits for additional epochs}
\label{app:pionier_fits}

\begin{figure*}[]
\centering
\includegraphics[width=\textwidth]{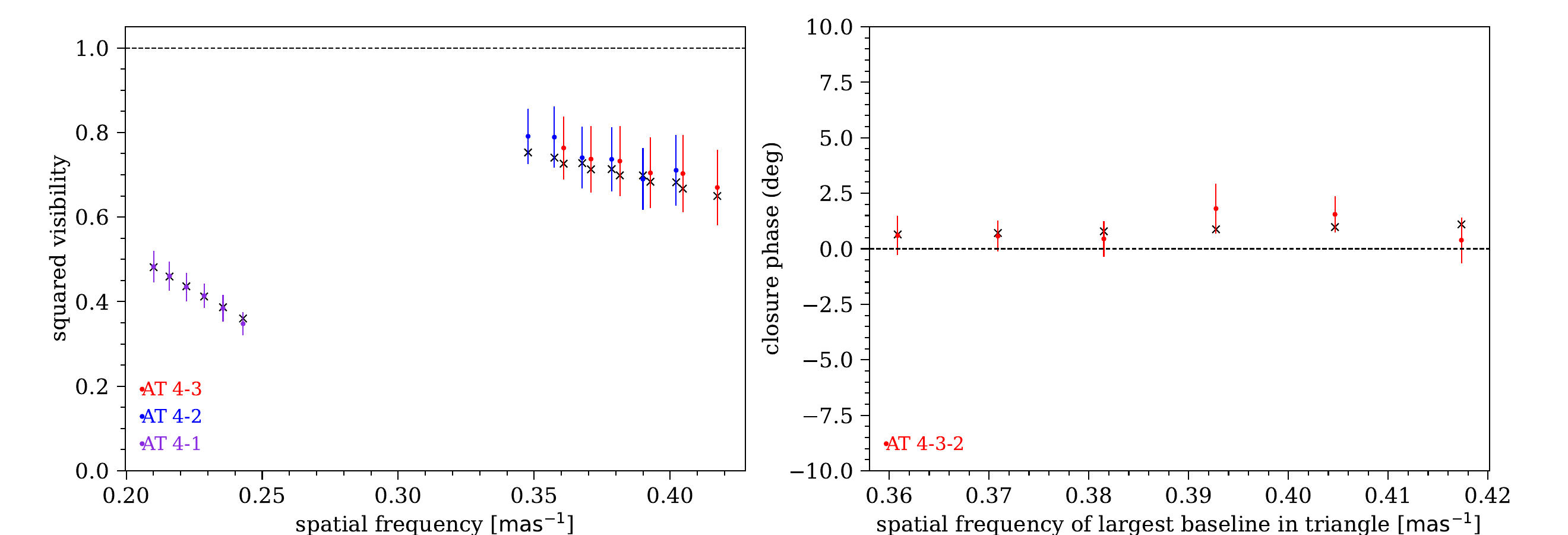} \\
\includegraphics[width=\textwidth]{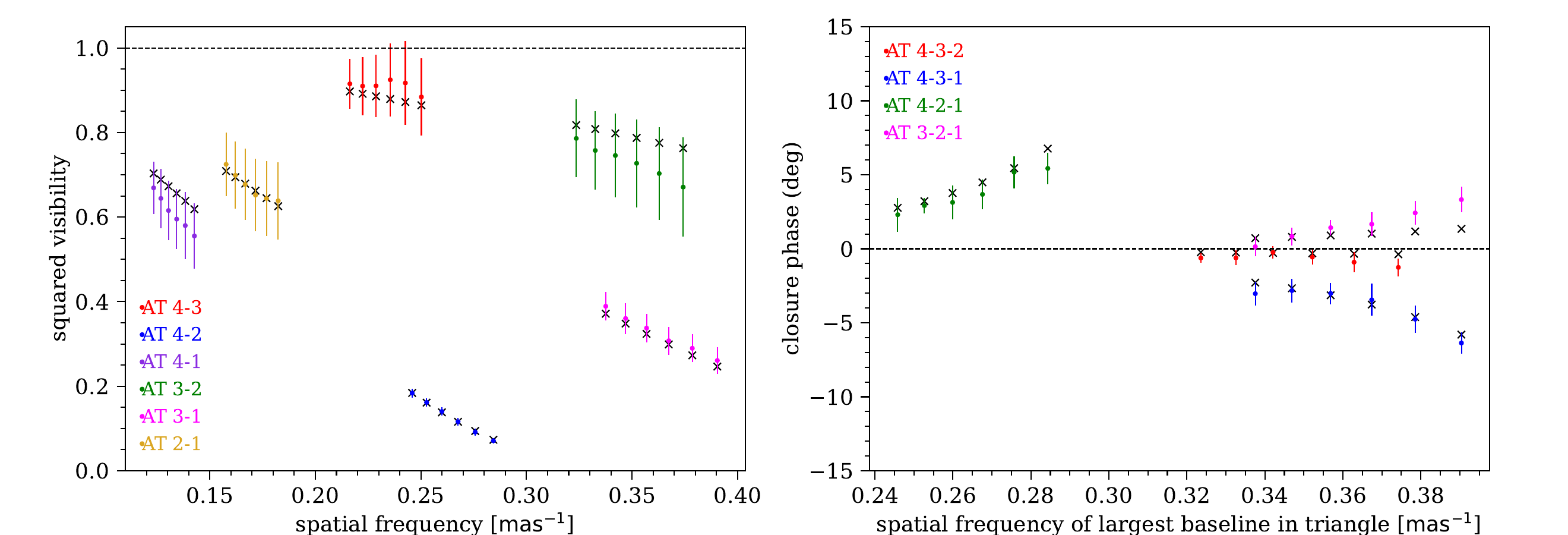} \\
\includegraphics[width=\textwidth]{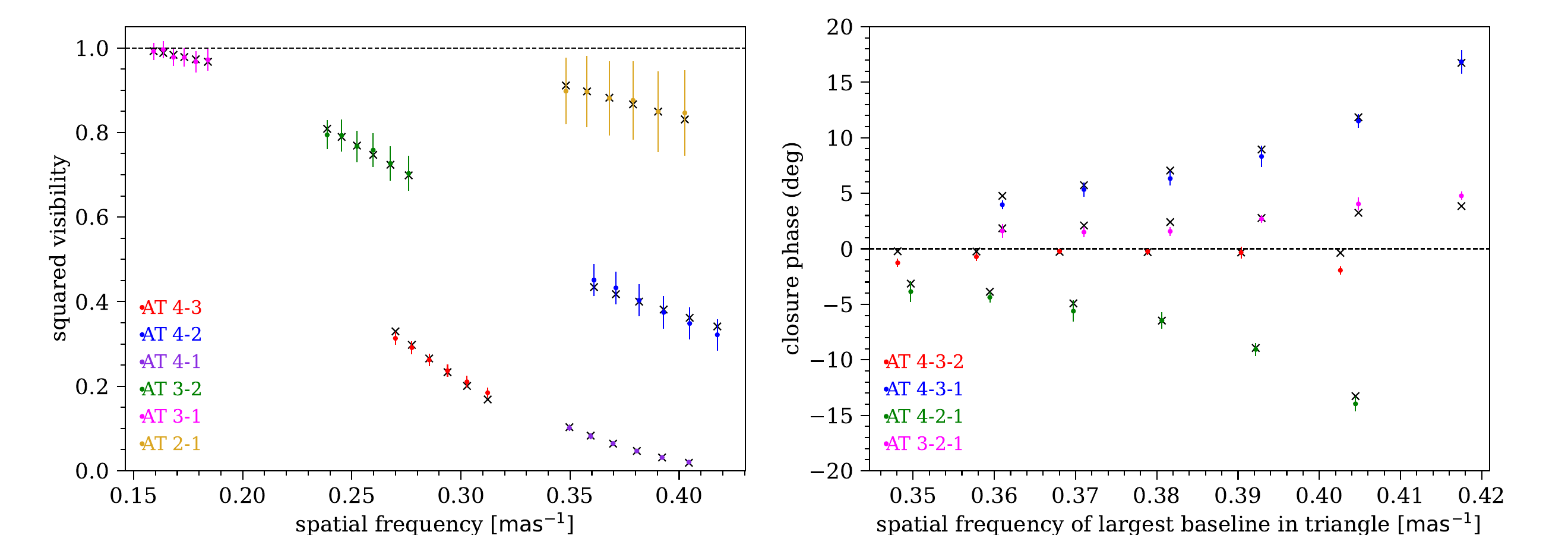} \\
\includegraphics[width=\textwidth]{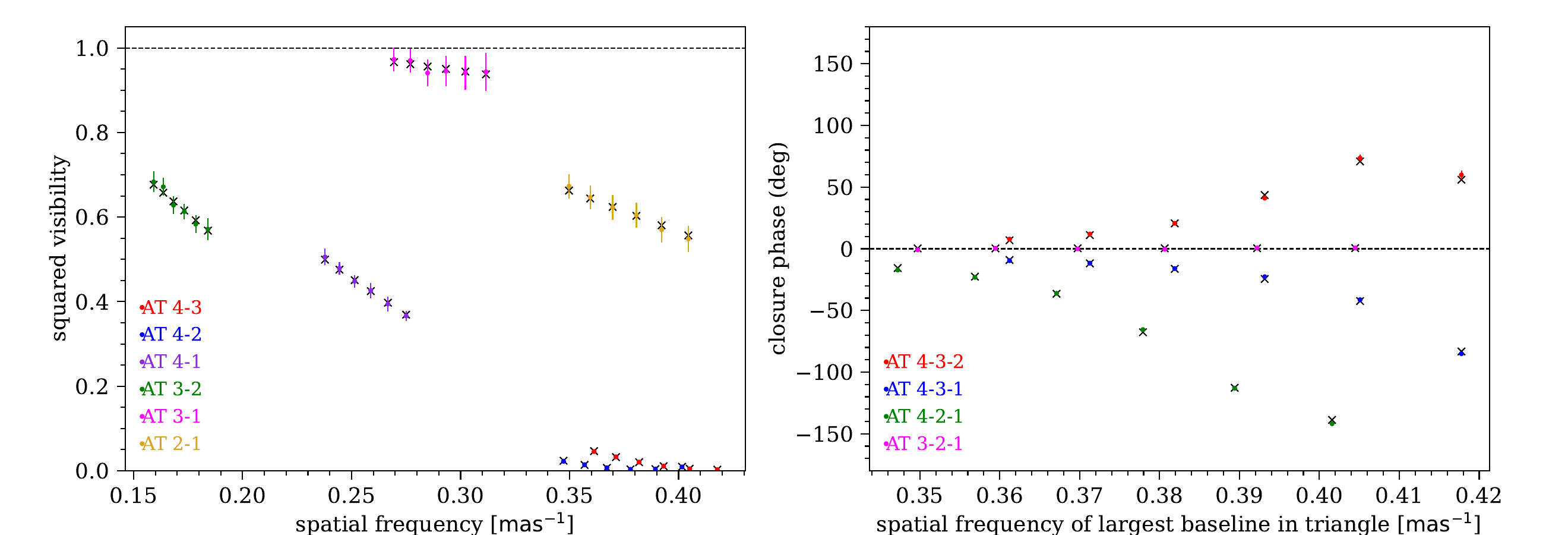}
\caption{\label{fig:pionier_fits_add} VLTI/PIONIER data (colored) and best fit binary model (solid black) for $\theta$ Ei A in epochs 2017-07-16, 2017-07-25, 2017-07-27 and 2017-07-28 (top to bottom). The dashed lines show the expected values for a single unresolved star.}
\end{figure*}

Figure \ref{fig:pionier_fits_add} shows the VLTI/PIONIER data (colored) and best fit binary model (solid black) for the remaining four epochs. 

\section{B. Full parameter distributions for orbital fit}
\label{app:corner_plot}

\begin{figure*}[]
\centering
\includegraphics[width=\textwidth]{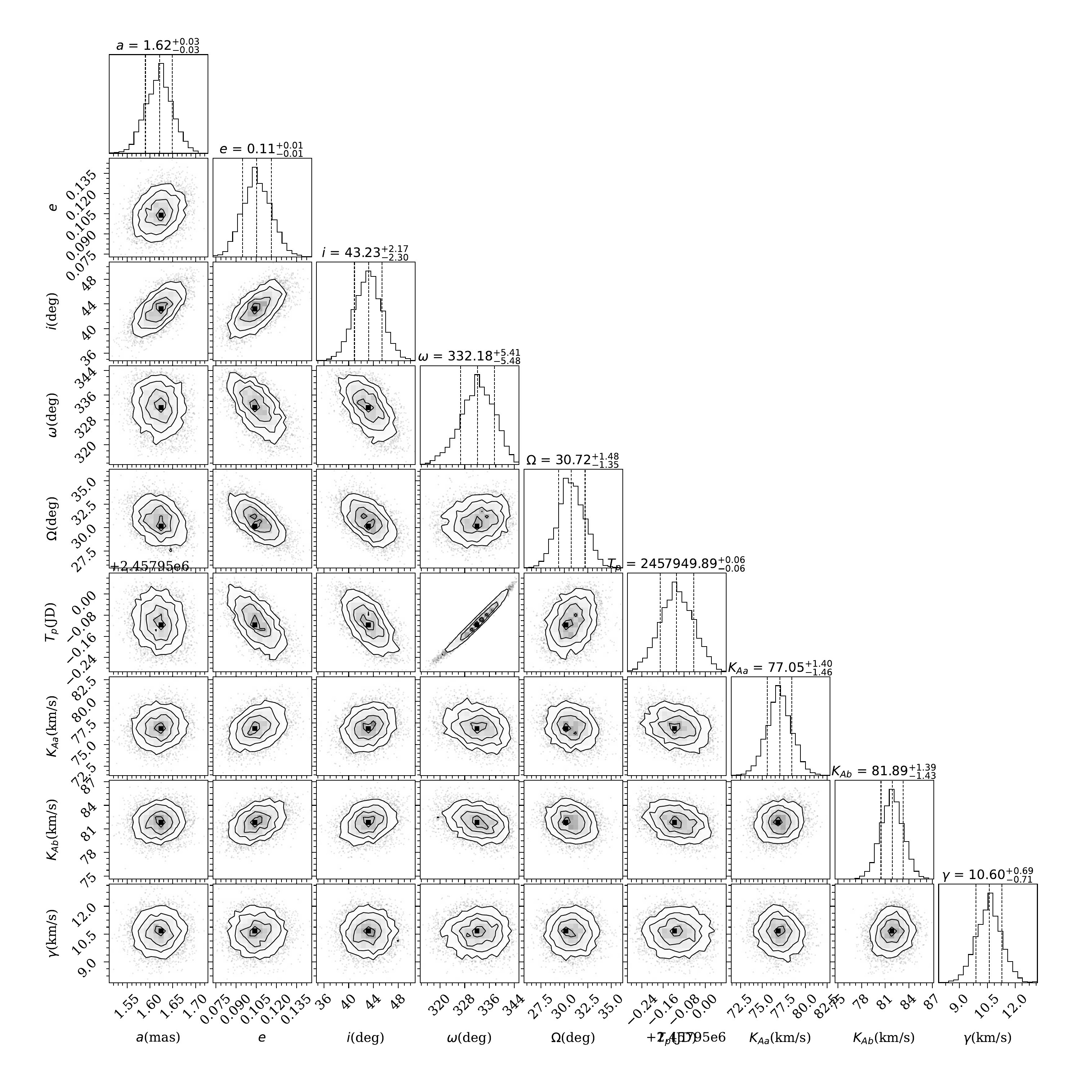}
\caption{\label{fig:corner} Orbital parameters distributions for $\theta$ Eri Aa+Ab.}
\end{figure*}

The full distributions of best fit orbital parameters for $\theta$ Eri Aa+Ab are shown in Figure \ref{fig:corner}. 

\section{C. Additional panels for the spectrum of Theta Eridani A for epoch 2015-09-19}
\label{app:spectrum_single}

\begin{figure*}[]
\includegraphics[width=\columnwidth]{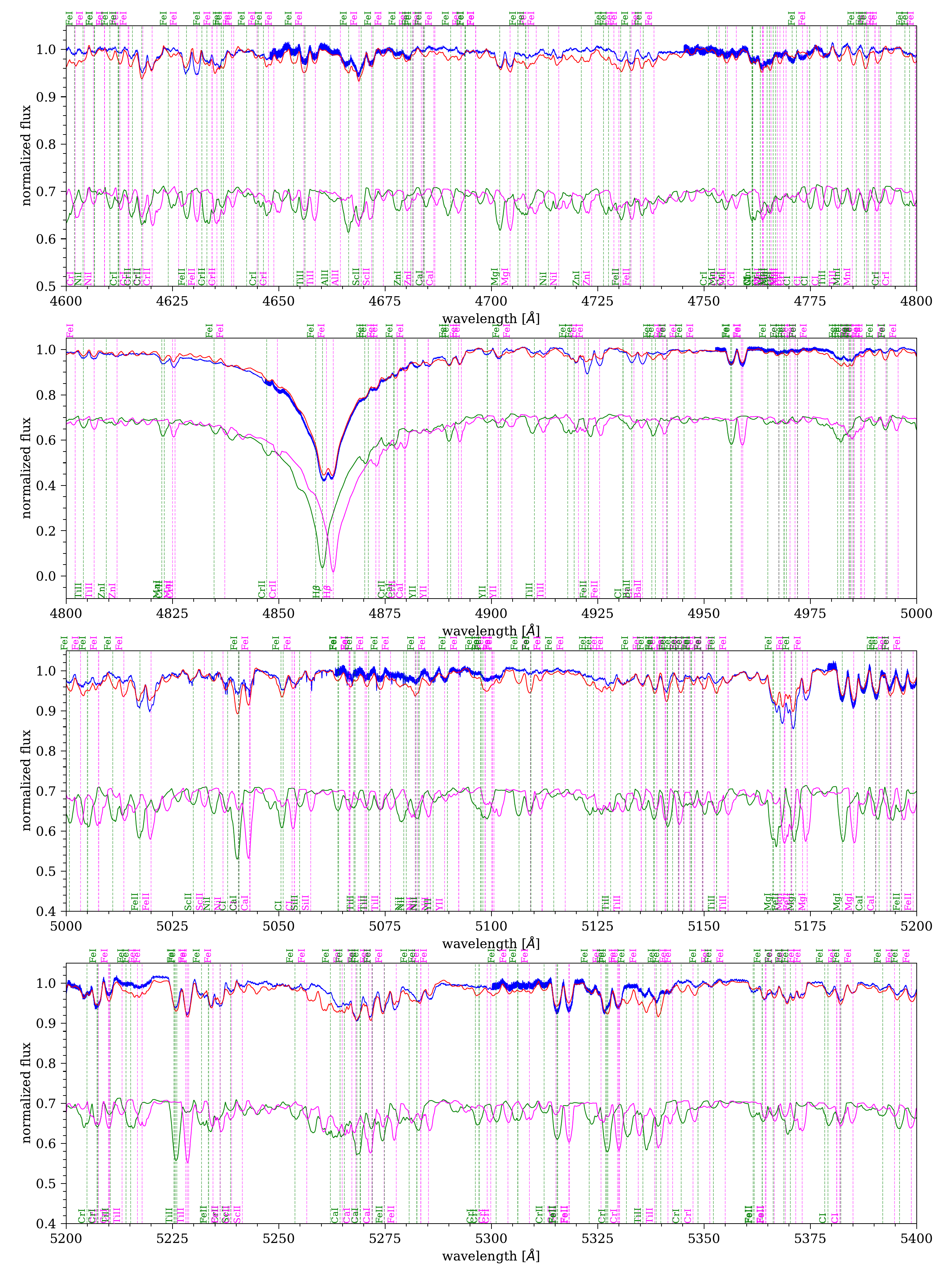}
\caption{Espadons normalized spectrum of Theta Eridani A (blue) for epoch 2015-09-19 (maximum velocity separation) together with normalized PHOENIX model spectra for the primary Aa (green), secondary Ab (magenta) and their sum (red). The magenta and green lines are shifted vertically by -0.3 for clarity.}
\label{fig:spectrum_2}
\end{figure*}

\begin{figure*}[]
\includegraphics[width=\columnwidth]{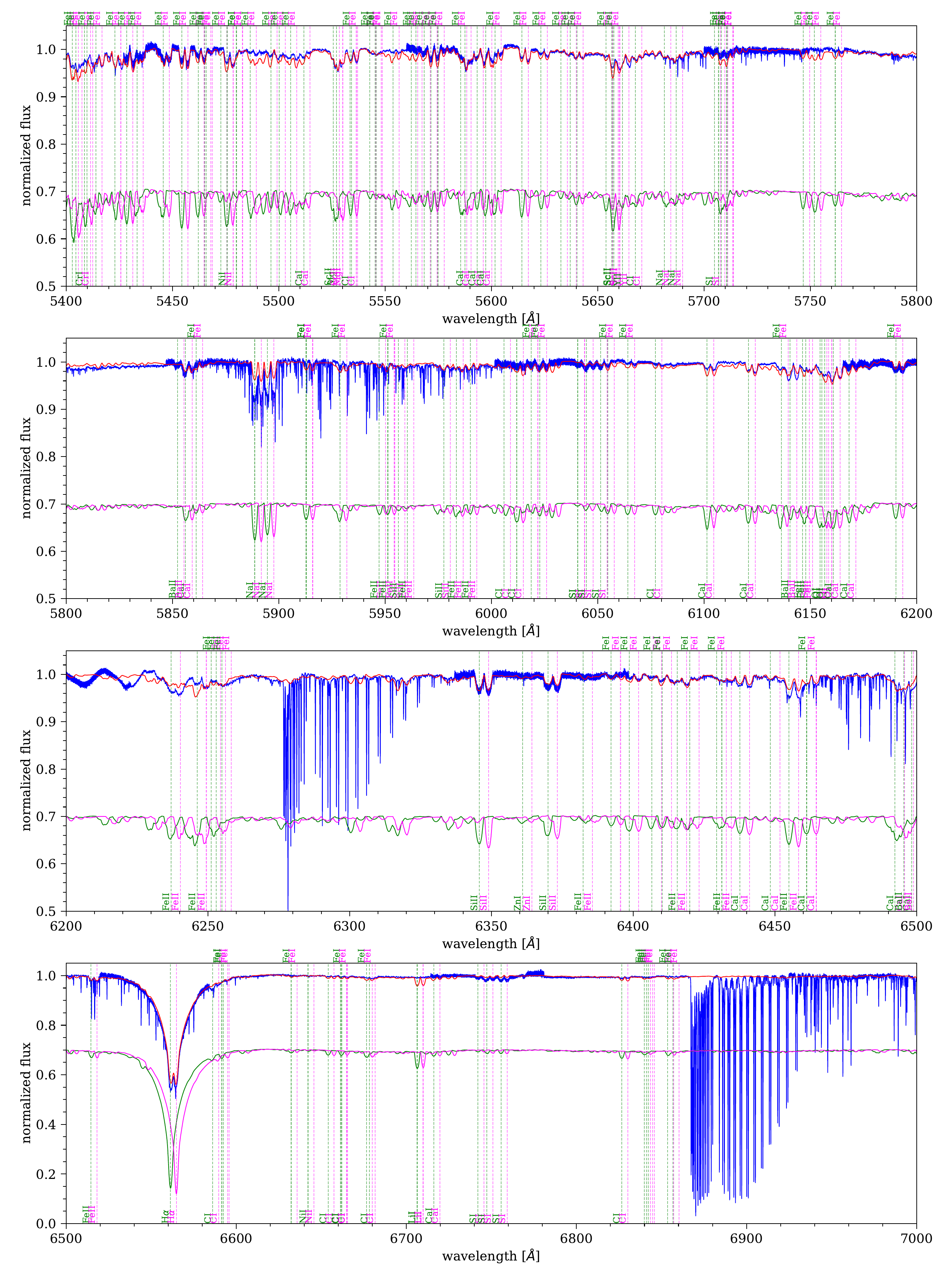}
\caption{Espadons normalized spectrum of Theta Eridani A (blue) for epoch 2015-09-19 (maximum velocity separation) together with normalized PHOENIX model spectra for the primary Aa (green), secondary Ab (magenta) and their sum (red). The magenta and green lines are shifted vertically by -0.3 for clarity.}
\label{fig:spectrum_3}
\end{figure*}

\begin{figure*}[]
\includegraphics[width=\columnwidth]{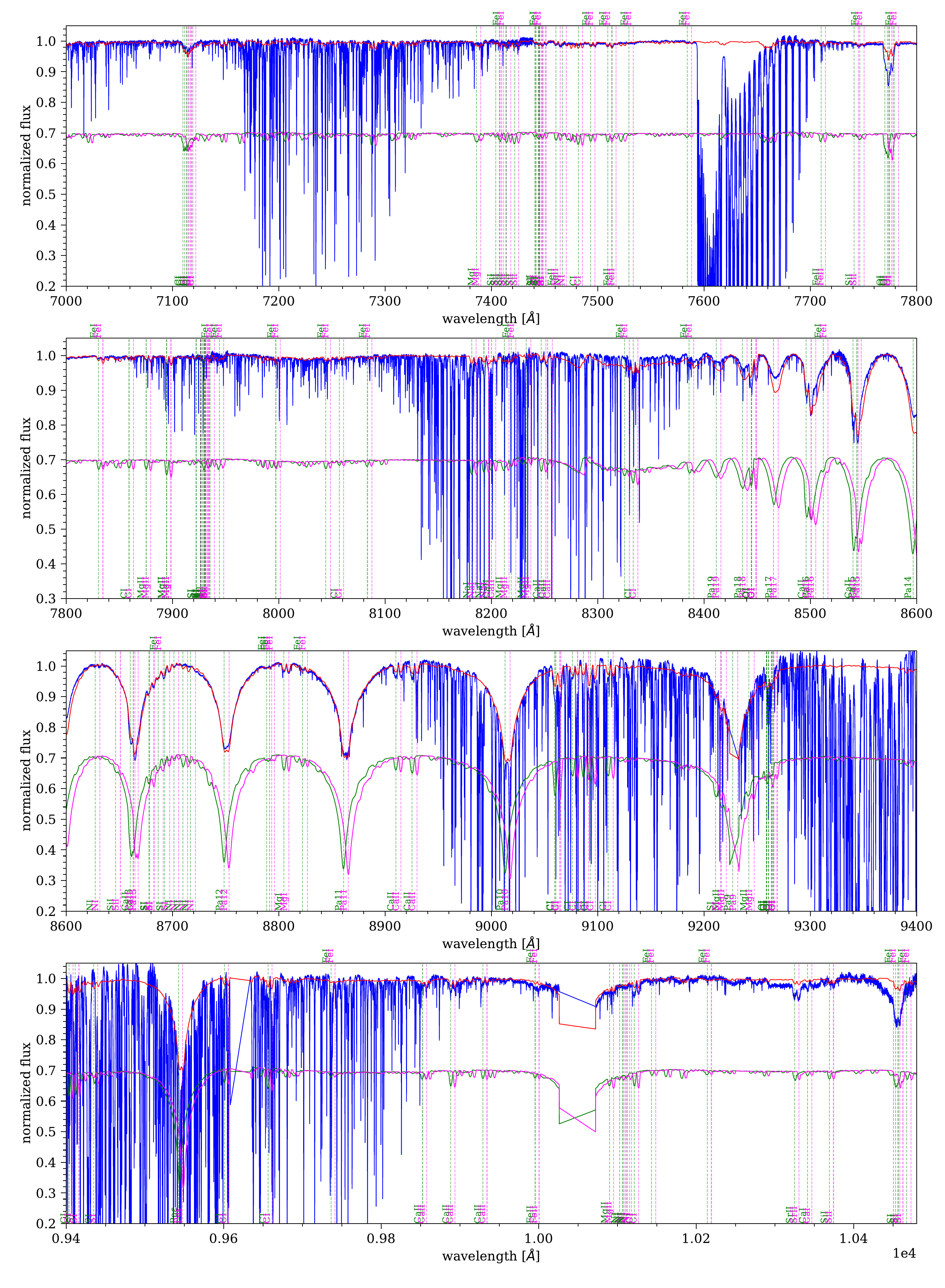}
\caption{Espadons normalized spectrum of Theta Eridani A (blue) for epoch 2015-09-19 (maximum velocity separation) together with normalized PHOENIX model spectra for the primary Aa (green), secondary Ab (magenta) and their sum (red). The magenta and green lines are shifted vertically by -0.3 for clarity.}
\label{fig:spectrum_4}
\end{figure*}

Figures \ref{fig:spectrum_2}, \ref{fig:spectrum_3}, \ref{fig:spectrum_4} show the remaining panels for the Espadons spectrum of Theta Eridani A for epoch 2015-09-19. 

\section{D. Spectrum of Theta Eridani A for epoch 2015-10-29}
\label{app:spectrum_single}

\begin{figure*}[]
\includegraphics[width=\columnwidth]{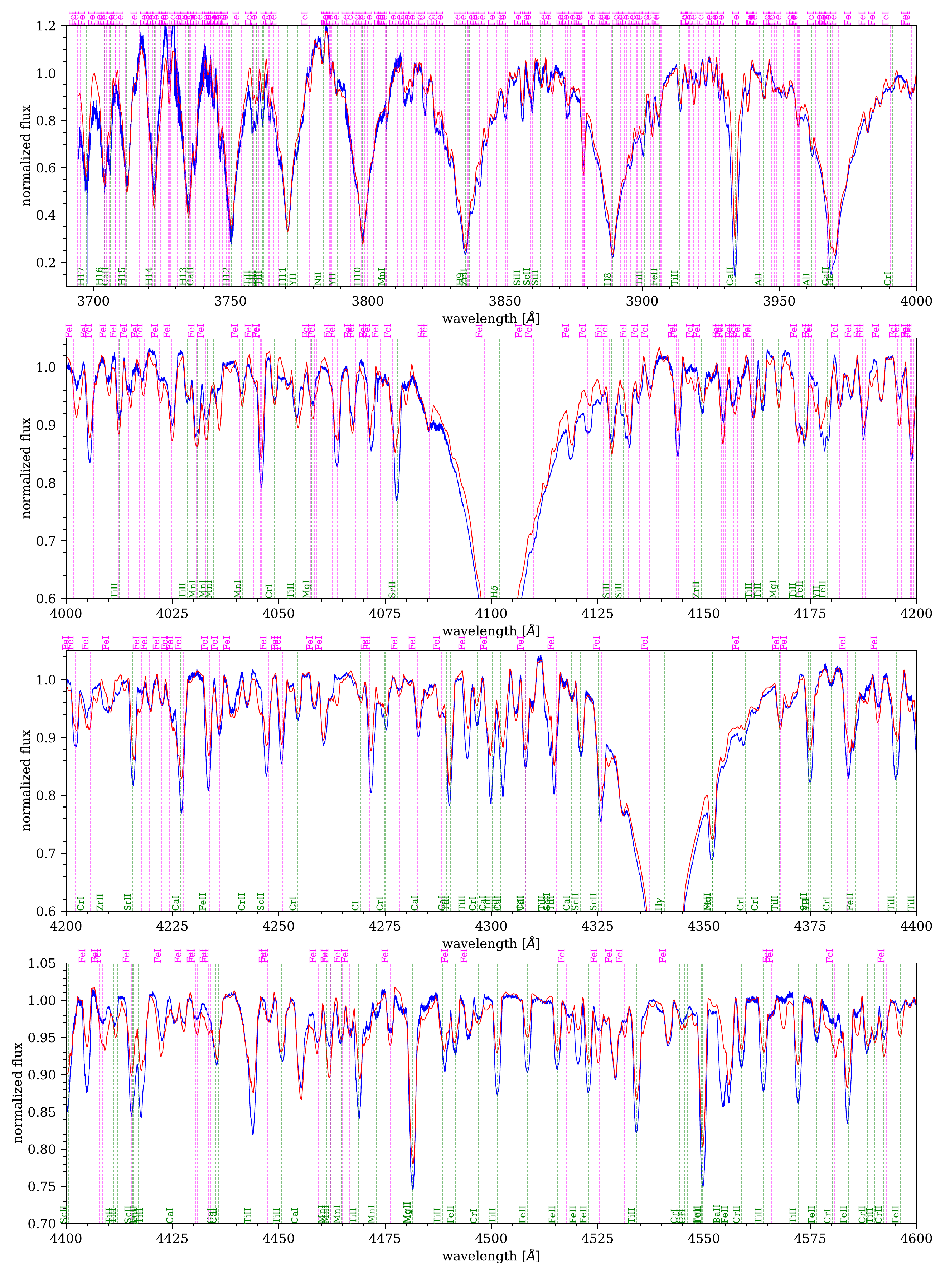}
\caption{Espadons normalized spectrum of Theta Eridani A (blue) for epoch 2015-10-29 (minimum velocity separation, so that the system appears as a single star) together with a PHOENIX model atmosphere (red) for the sum of Aa and Ab.}
\label{fig:spectrum_single1}
\end{figure*}

\begin{figure*}[]
\includegraphics[width=\columnwidth]{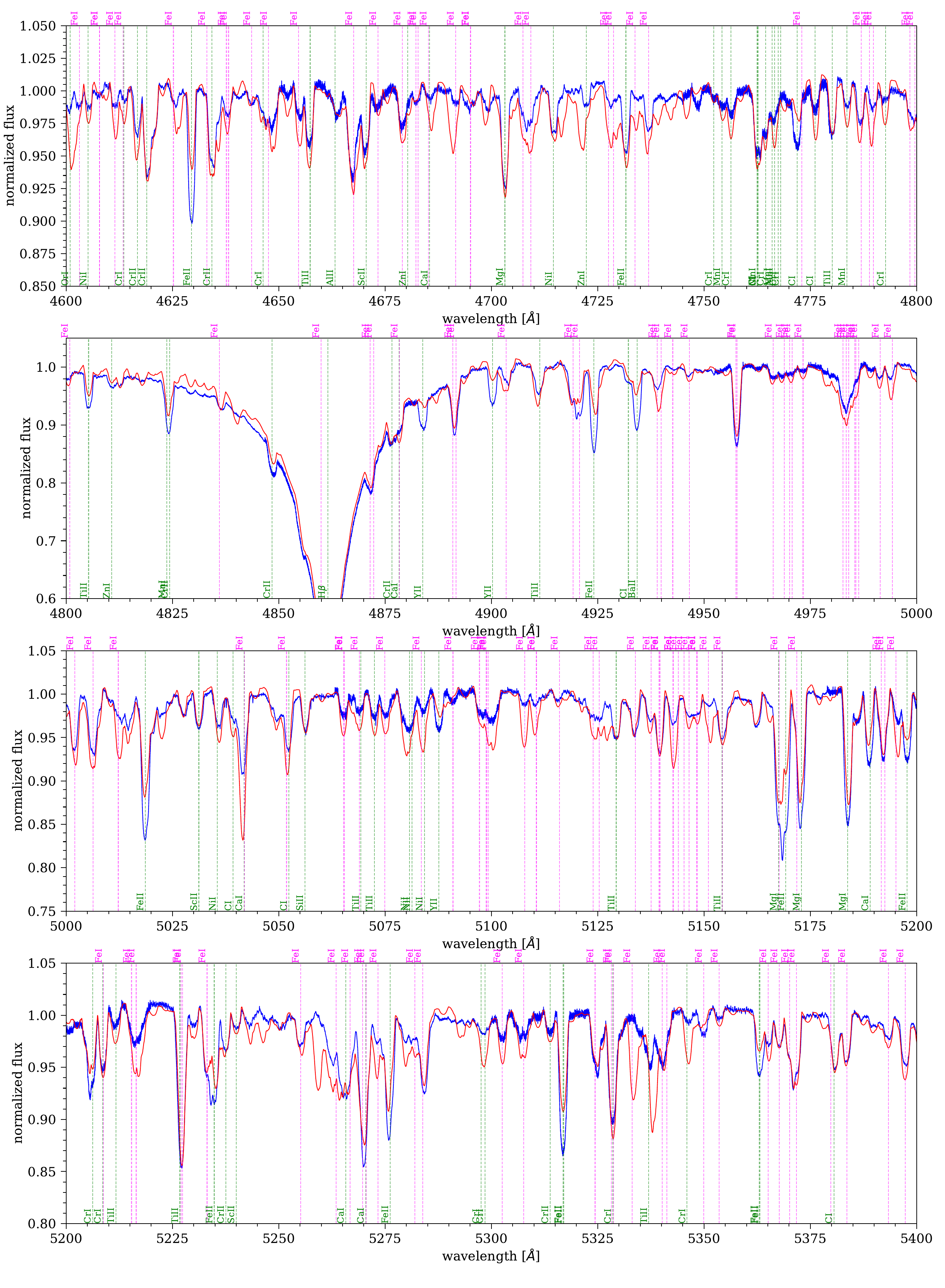}
\caption{Espadons normalized spectrum of Theta Eridani A (blue) for epoch 2015-10-29 together with a PHOENIX model atmosphere (red) for the sum of Aa and Ab.}
\label{fig:spectrum_single2}
\end{figure*}

\begin{figure*}[]
\includegraphics[width=\columnwidth]{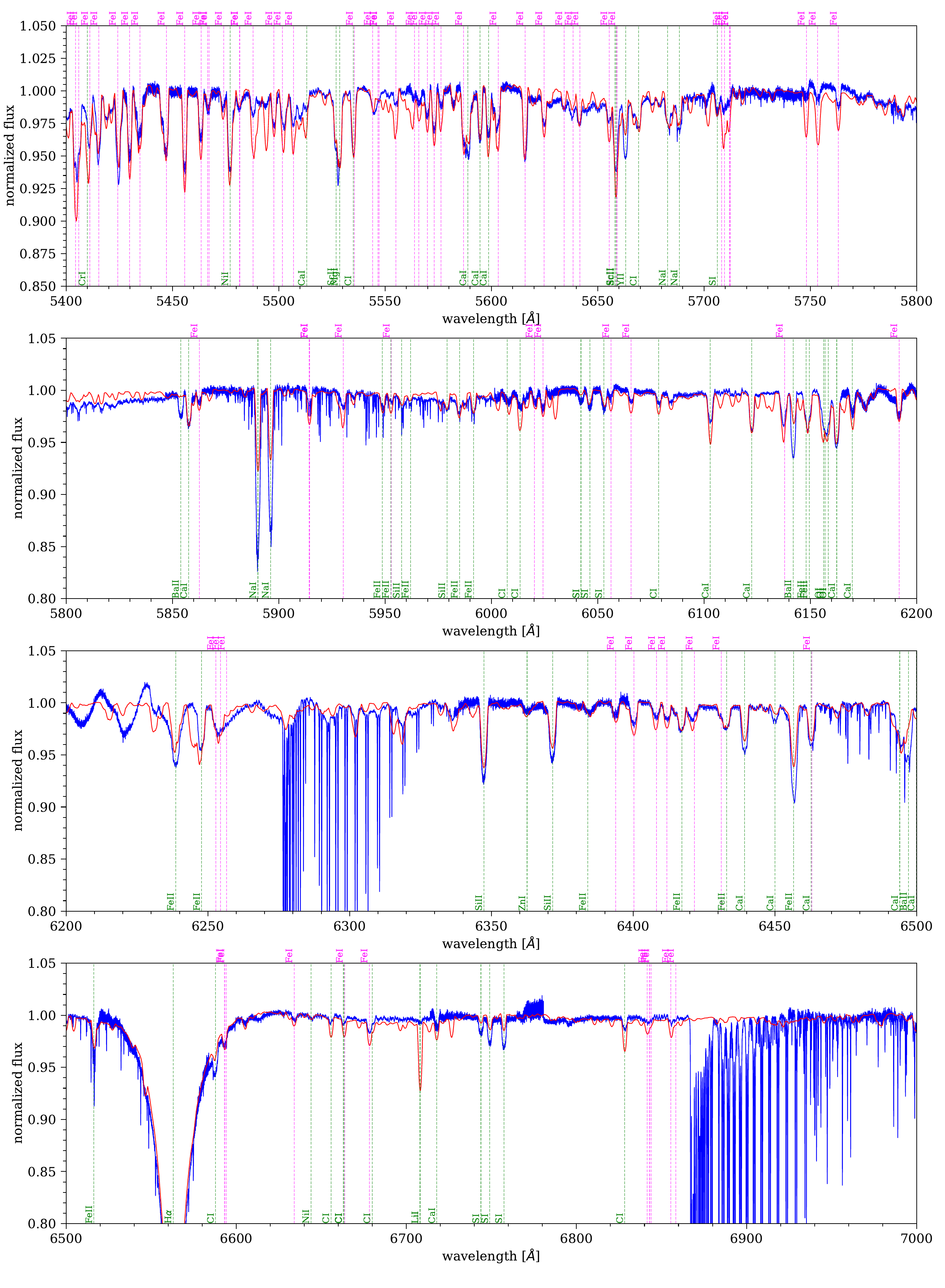}
\caption{Espadons normalized spectrum of Theta Eridani A (blue) for epoch 2015-10-29 together with a PHOENIX model atmosphere (red) for the sum of Aa and Ab.}
\label{fig:spectrum_single3}
\end{figure*}

\begin{figure*}[]
\includegraphics[width=\columnwidth]{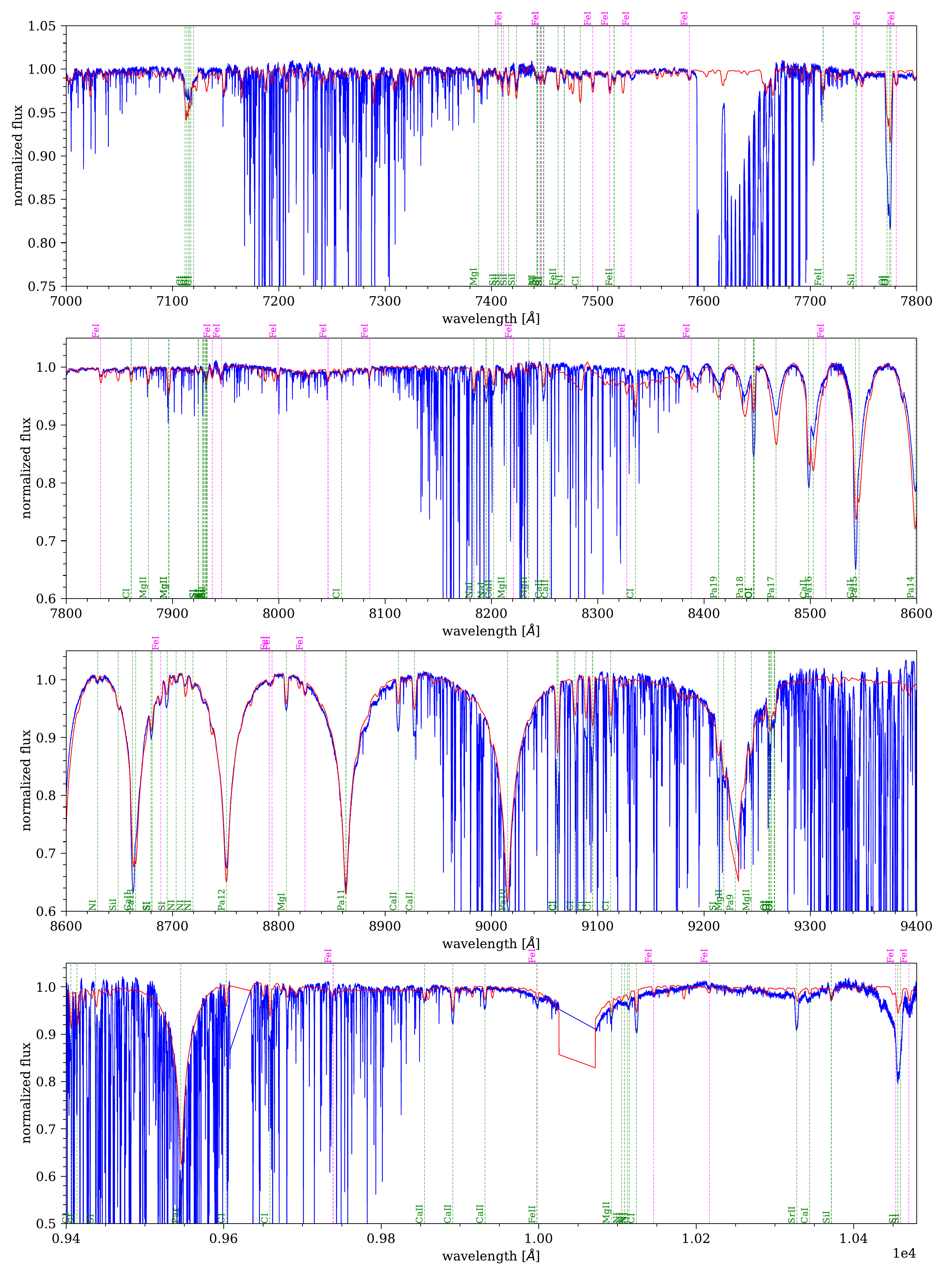}
\caption{Espadons normalized spectrum of Theta Eridani A (blue) for epoch 2015-10-29 together with a PHOENIX model atmosphere (red) for the sum of Aa and Ab.}
\label{fig:spectrum_single4}
\end{figure*}

Figures \ref{fig:spectrum_single1}, \ref{fig:spectrum_single2}, \ref{fig:spectrum_single3} and \ref{fig:spectrum_single4} show the Espadons normalized spectrum of Theta Eridani A (blue) for epoch 2015-10-29, in which Aa and Ab have the minimum velocity separation so that their spectra are blended into an apparently single star. The sum of the PHOENIX model atmospheres for Aa and Ab is shown in red. Line identifications are marked in magenta for Fe I and in green for all remaining species. 

\section{E. Spectrum of Theta Eridani B}
\label{app:spectrum_single}

\begin{figure*}[]
\includegraphics[width=\columnwidth]{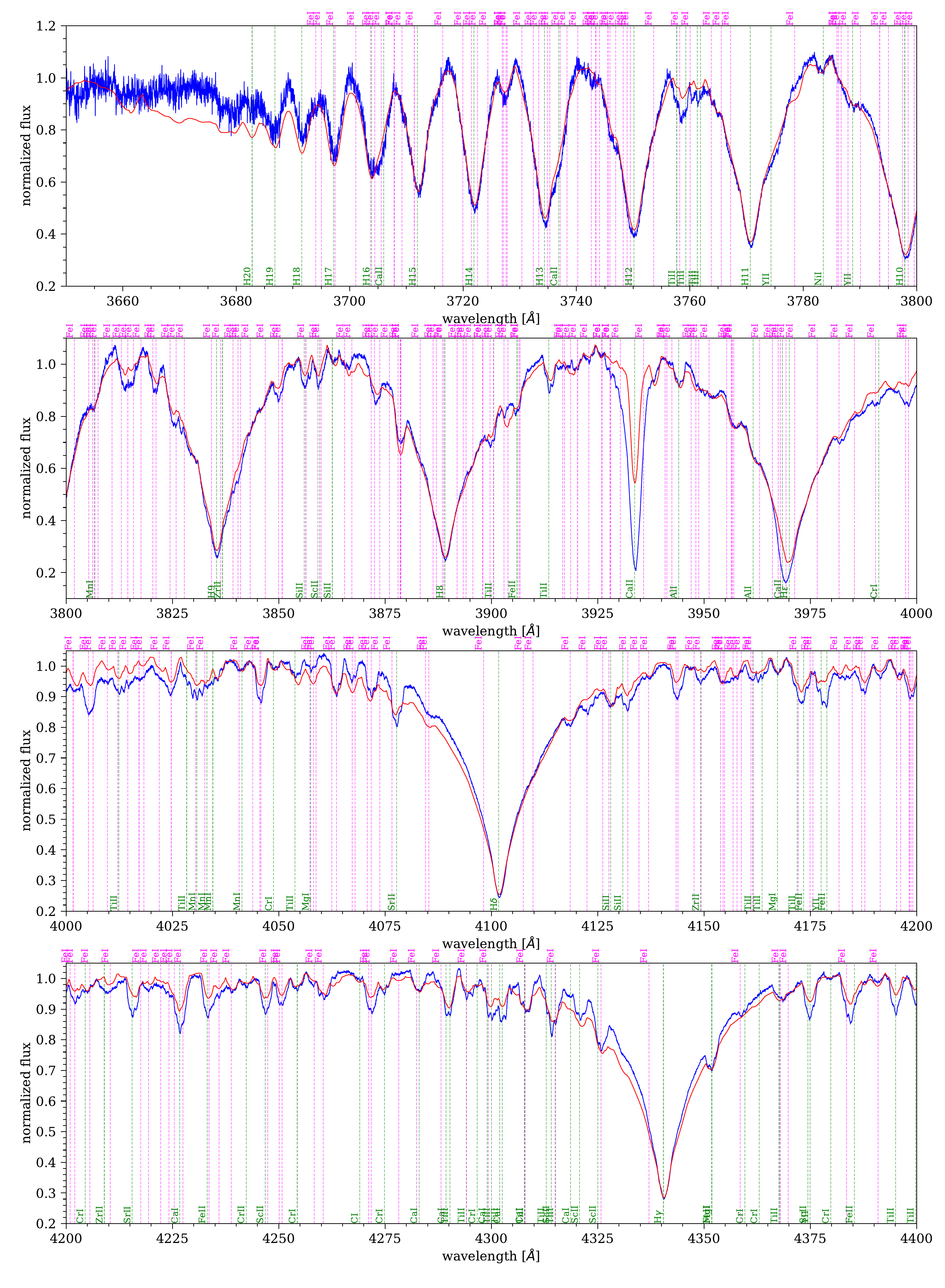}
\caption{FEROS normalized spectrum of Theta Eridani B (blue) and a corresponding PHOENIX model atmosphere (red).}
\label{fig:spectrum_B1}
\end{figure*}

\begin{figure*}[]
\includegraphics[width=\columnwidth]{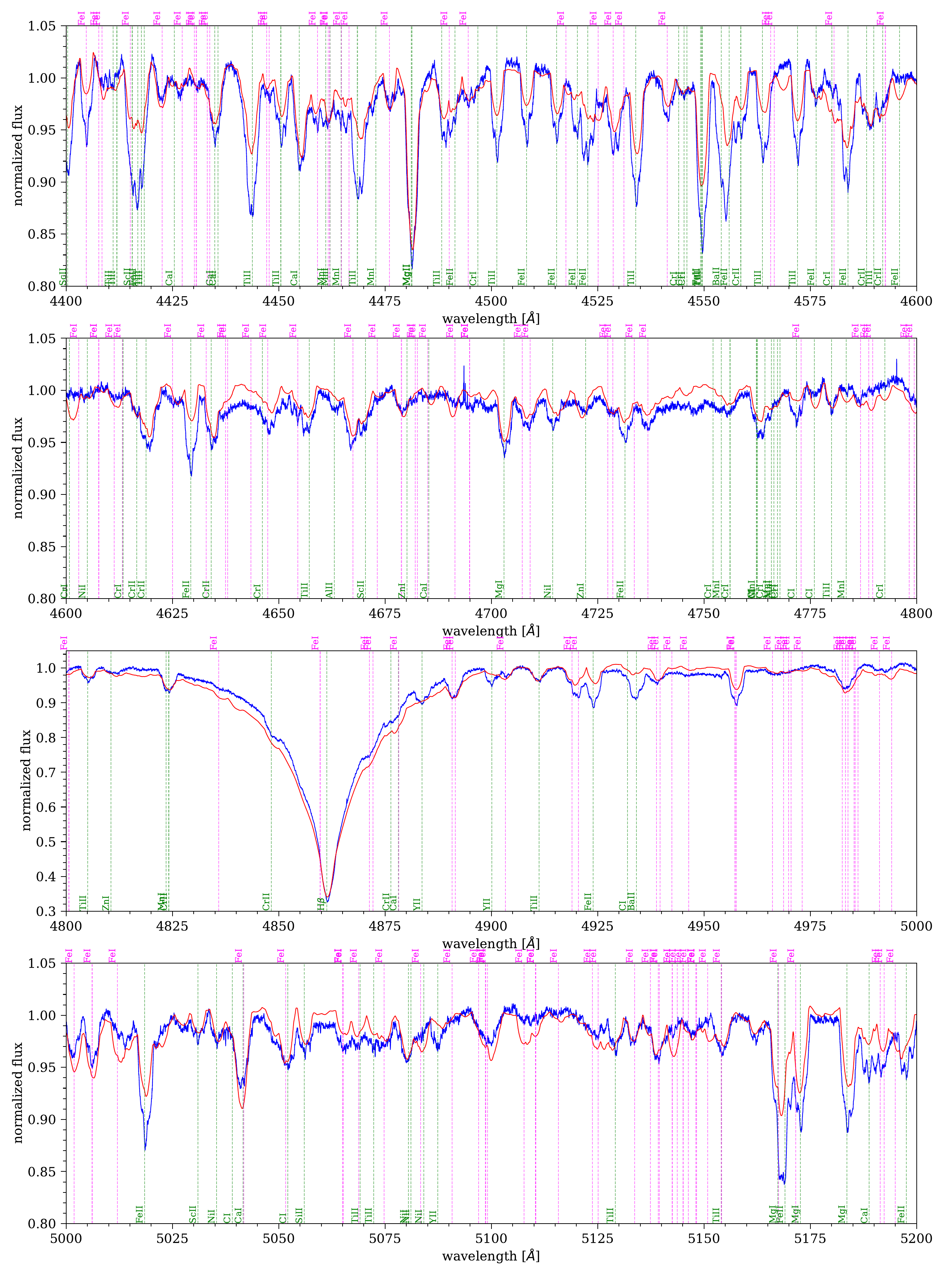}
\caption{FEROS normalized spectrum of Theta Eridani B (blue) and a corresponding PHOENIX model atmosphere (red).}
\label{fig:spectrum_B2}
\end{figure*}

\begin{figure*}[]
\includegraphics[width=\columnwidth]{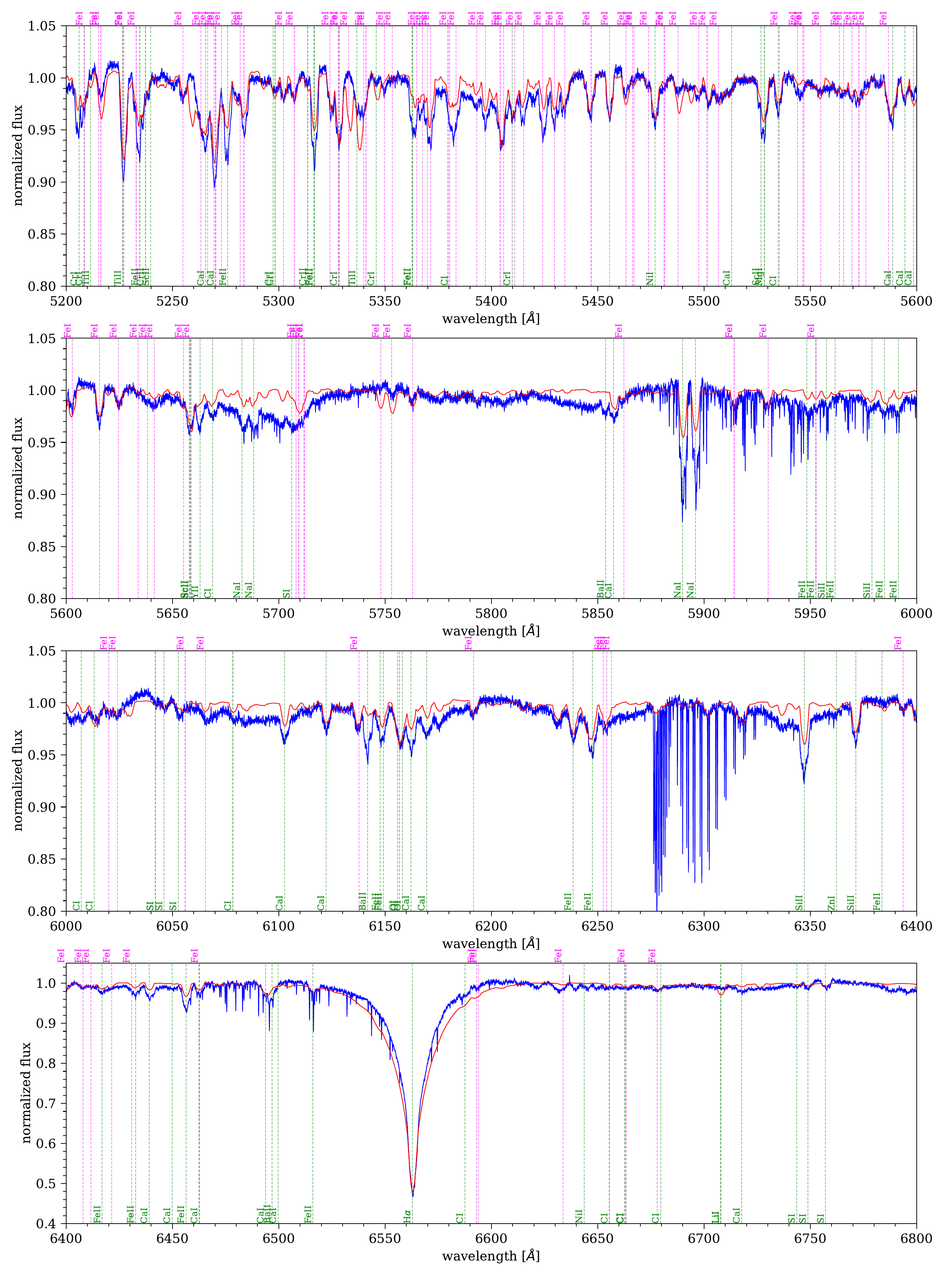}
\caption{FEROS normalized spectrum of Theta Eridani B (blue) and a corresponding PHOENIX model atmosphere (red).}
\label{fig:spectrum_B3}
\end{figure*}

\begin{figure*}[]
\includegraphics[width=\columnwidth]{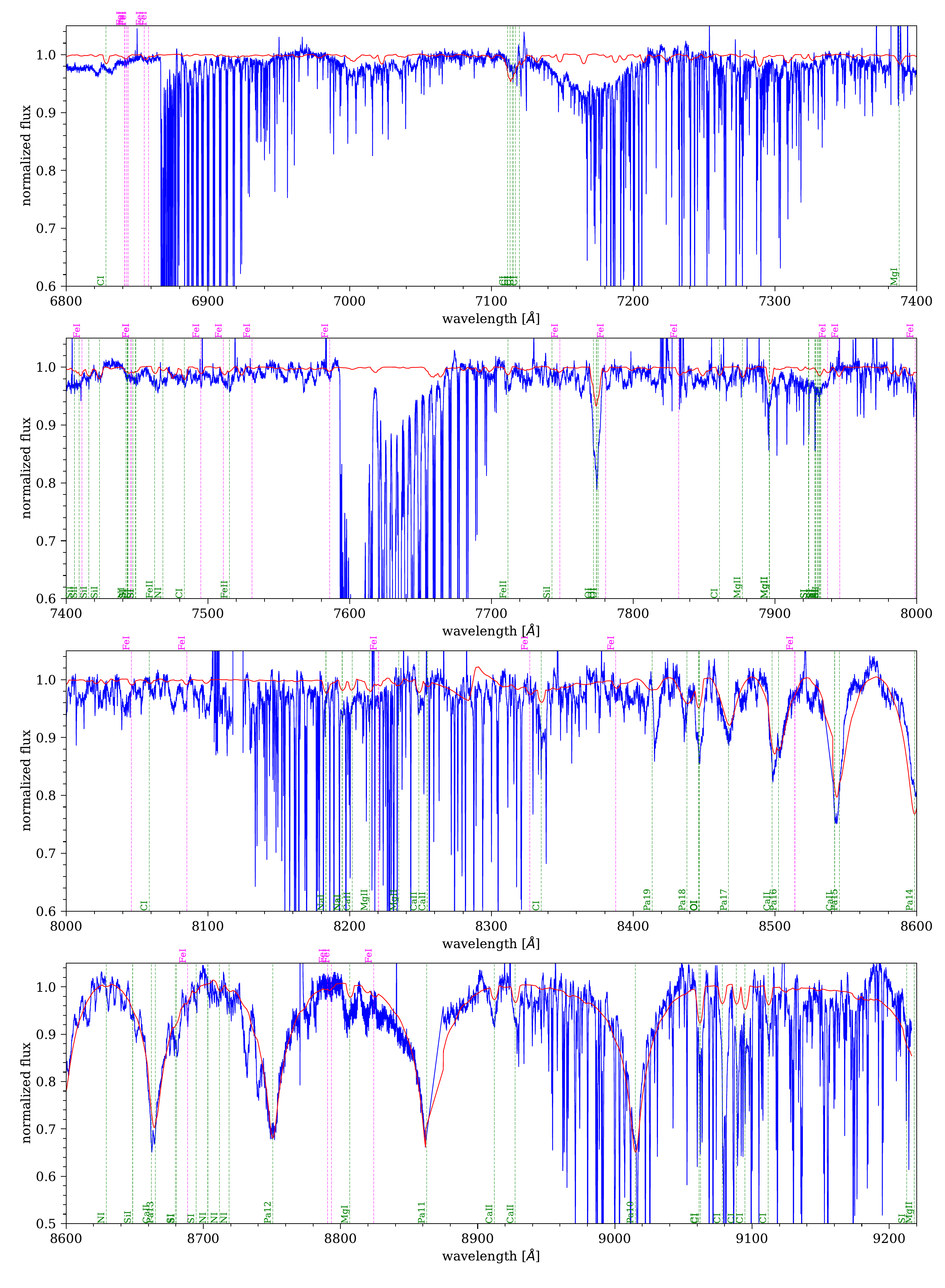}
\caption{FEROS normalized spectrum of Theta Eridani B (blue) and a corresponding PHOENIX model atmosphere (red).}
\label{fig:spectrum_B4}
\end{figure*}

Figures \ref{fig:spectrum_B1}, \ref{fig:spectrum_B2}, \ref{fig:spectrum_B3} and \ref{fig:spectrum_B4} show the normalized FEROS spectrum of Theta Eridani B (blue) together with a PHOENIX model atmosphere with $\log g=4.0$ and $T=8200 \text{ K}$. 

\section{F. Lines identified in the spectrum of Theta Eridani A.}
\label{app:line_list}

Table \ref{table:line_list} lists the lines identified in the spectrum of Theta Eridani A. 

\clearpage

\renewcommand{\arraystretch}{1.4}
\begin{longtable}{cccccccc}
\caption{\label{table:line_list} Identified spectral lines in Theta Eridani A.} \\
\toprule
Ion & Wavelength (\AA) & Ion & Wavelength (\AA) & Ion & Wavelength (\AA) & Ion & Wavelength (\AA) \\
\midrule
\endfirsthead
\toprule
Ion & Wavelength (\AA) & Ion & Wavelength (\AA) & Ion & Wavelength (\AA) & Ion & Wavelength (\AA) \\
\midrule
\endhead
\bottomrule
\endfoot
\midrule
H17 & 3697.15 & H16 & 3703.85 & H15 & 3711.97 & H14 & 3721.94 \\[2pt]
H13 & 3734.37 & H12 & 3750.15 & H11 & 3770.63 & H10 & 3797.90 \\[2pt]
H9 & 3835.38 & H8 & 3889.05 & H$\epsilon$ & 3970.07 & H$\delta$ & 4101.735 \\[2pt]
H$\gamma$ & 4340.463 & H$\beta$ & 4861.325 & H$\alpha$ & 6562.80 & Pa19 & 8413.38 \\[2pt]
Pa18 & 8438.02 & Pa17 & 8467.31 & Pa16 & 8502.54 & Pa15 & 8545.44 \\[2pt]
Pa14 & 8598.46 & Pa13 & 8665.08 & Pa12 & 8750.54 & Pa11 & 8862.85 \\[2pt]
Pa10 & 9014.9 & Pa9 & 9229.0 & Pa$\epsilon$ & 9545.0 &  &\\[2pt]
\midrule
LiI & 6707.76 & LiI & 6707.91 &  &  &  &  \\[2pt]
\midrule
CI & 4269.0197 & CI & 4762.3033 & CI & 4762.5252 & CI & 4771.73346 \\[2pt]
CI & 4775.907 & CI & 4932.02553 & CI & 5039.07 & CI & 5052.149 \\[2pt]
CI & 5380.3308 & CI & 5534.807 & CI & 5668.951 & CI & 6007.178 \\[2pt]
CI & 6013.22 & CI & 6078.395 & CI & 6587.608 & CI & 6655.509 \\[2pt]
CI & 6662.733 & CI & 6663.044 & CI & 6679.649 & CI & 6828.117 \\[2pt]
CI & 7111.475 & CI & 7113.180 & CI & 7115.19 & CI & 7115.19 \\[2pt]
CI & 7116.990 & CI & 7119.671 & CI & 7483.41 & CI & 7860.88 \\[2pt]
CI & 8058.59 & CI & 8335.1443 & CI & 9061.4347 & CI & 9062.4723 \\[2pt]
CI & 9078.2819 & CI & 9088.5097 & CI & 9094.8303 & CI & 9111.8016 \\[2pt]
CI & 9405.7281 & CI & 9603.0309 & CI & 9658.4377 & CI & 10123.860 \\[2pt]
\midrule
NI & 7442.29 & NI & 7468.31 & NI & 8629.24 & NI & 8703.25 \\[2pt]
NI & 8711.70 & NI & 8718.83 & NI & 10105.13 & NI & 10108.89 \\[2pt]
NI & 10112.48 & NI & 10114.64 &  &  &  &  \\[2pt]
\midrule
OI & 6155.98 & OI & 6156.77 & OI & 6158.18 & OI & 7771.94 \\[2pt]
OI & 7774.17 & OI & 7775.39 & OI & 8446.25 & OI & 8446.36 \\[2pt]
OI & 8446.76 & OI & 9260.81 & OI & 9260.84 & OI & 9260.94 \\[2pt]
OI & 9262.58 & OI & 9262.67 & OI & 9262.77 & OI & 9265.826 \\[2pt]
OI & 9265.94 & OI & 9266.01 &  &  &  &  \\[2pt]
\midrule
NaI & 5682.6333 & NaI & 5688.2046 & NaI & 5889.95095 & NaI & 5895.92424 \\[2pt]
NaI & 8183.2556 & NaI & 8194.7905 & NaI & 8194.8237 &  &  \\[2pt]
\midrule
MgI & 4057.5052 & MgI & 4167.2712 & MgI & 4351.9056 & MgI & 4702.9909 \\[2pt]
MgI & 5167.3216 & MgI & 5172.6843 & MgI & 5183.6042 & MgI & 5528.4047 \\[2pt]
MgI & 7387.685 & MgI & 8806.757 &  &  &  &  \\[2pt]
\midrule
MgII & 4481.130 & MgII & 4481.327 & MgII & 7877.051 & MgII & 7896.04 \\[2pt]
MgII & 7896.368 & MgII & 8213.989 & MgII & 8234.639 & MgII & 9218.248 \\[2pt]
MgII & 9244.266 & MgII & 10092.160 &  &  &  &  \\[2pt]
\midrule
AlI & 3944.0058 & AlI & 3961.5200 &  &  &  &  \\[2pt]
\midrule
AlII & 4663.056 &  &  &  &  &  &  \\[2pt]
\midrule
SiI & 7405.774 & SiI & 7409.082 & SiI & 7415.946 & SiI & 7423.497 \\[2pt]
SiI & 7742.71 & SiI & 7932.349 & SiI & 8648.462 & SiI & 10371.264 \\[2pt]
\midrule
SiII & 3856.02 & SiII & 3862.60 & SiII & 4128.07 & SiII & 4130.89 \\[2pt]
SiII & 5055.98 & SiII & 5957.56 & SiII & 5978.93 & SiII & 6347.10 \\[2pt]
SiII & 6371.36 &  &  &  &  &  &  \\[2pt]
\midrule
SI & 5706.11 & SI & 6041.93 & SI & 6041.93 & SI & 6046.04 \\[2pt]
SI & 6052.66 & SI & 6743.58 & SI & 6748.79 & SI & 6757.16 \\[2pt]
SI & 7443.35 & SI & 7443.35 & SI & 7443.35 & SI & 7446.97 \\[2pt]
SI & 7449.02 & SI & 7449.02 & SI & 7449.02 & SI & 7923.90 \\[2pt]
SI & 7923.90 & SI & 7923.90 & SI & 7927.98 & SI & 7928.82 \\[2pt]
SI & 7928.82 & SI & 7930.28 & SI & 7930.28 & SI & 7930.28 \\[2pt]
SI & 7931.66 & SI & 7931.66 & SI & 7931.66 & SI & 8679.65 \\[2pt]
SI & 8680.46 & SI & 8694.71 & SI & 9212.851 & SI & 9413.382 \\[2pt]
SI & 9437.072 & SI & 10455.441 & SI & 10459.402 &  &  \\[2pt]
\midrule
CaI & 4226.73 & CaI & 4283.01 & CaI & 4289.36 & CaI & 4298.99 \\[2pt]
CaI & 4302.53 & CaI & 4307.74 & CaI & 4318.65 & CaI & 4425.44 \\[2pt]
CaI & 4434.96 & CaI & 4435.69 & CaI & 4454.78 & CaI & 4685.27 \\[2pt]
CaI & 4878.13 & CaI & 5041.62 & CaI & 5188.85 & CaI & 5265.56 \\[2pt]
CaI & 5270.27 & CaI & 5512.98 & CaI & 5588.76 & CaI & 5594.47 \\[2pt]
CaI & 5598.49 & CaI & 5857.45 & CaI & 6102.72 & CaI & 6122.22 \\[2pt]
CaI & 6162.17 & CaI & 6169.56 & CaI & 6439.07 & CaI & 6449.81 \\[2pt]
CaI & 6462.57 & CaI & 6493.78 & CaI & 6499.65 & CaI & 6717.68 \\[2pt]
CaI & 10343.81 &  &  &  &  &  &  \\[2pt]
\midrule
CaII & 3706.03 & CaII & 3736.90 & CaII & 3933.66 & CaII & 3968.47 \\[2pt]
CaII & 8201.72 & CaII & 8248.80 & CaII & 8254.73 & CaII & 8498.02 \\[2pt]
CaII & 8542.09 & CaII & 8662.14 & CaII & 8912.07 & CaII & 8927.36 \\[2pt]
CaII & 9854.76 & CaII & 9890.63 & CaII & 9931.39 &  &  \\[2pt]
\midrule
ScII & 3859.592 & ScII & 4246.820 & ScII & 4314.082 & ScII & 4320.745 \\[2pt]
ScII & 4324.998 & ScII & 4374.462 & ScII & 4400.386 & ScII & 4415.544 \\[2pt]
ScII & 4670.406 & ScII & 5031.010 & ScII & 5239.811 & ScII & 5526.785 \\[2pt]
ScII & 5657.907 & ScII & 5658.362 &  &  &  &  \\[2pt]
\midrule
TiII & 3757.6850 & TiII & 3759.2915 & TiII & 3761.3202 & TiII & 3761.8721 \\[2pt]
TiII & 3900.5389 & TiII & 3913.4614 & TiII & 4012.3836 & TiII & 4028.3384 \\[2pt]
TiII & 4053.8210 & TiII & 4161.5293 & TiII & 4163.6437 & TiII & 4171.9038 \\[2pt]
TiII & 4290.2148 & TiII & 4294.0939 & TiII & 4300.0421 & TiII & 4301.9225 \\[2pt]
TiII & 4307.8657 & TiII & 4312.8600 & TiII & 4314.9708 & TiII & 4367.6521 \\[2pt]
TiII & 4395.0312 & TiII & 4399.7652 & TiII & 4411.0724 & TiII & 4411.929 \\[2pt]
TiII & 4417.7137 & TiII & 4418.3313 & TiII & 4443.8007 & TiII & 4450.4822 \\[2pt]
TiII & 4468.4924 & TiII & 4488.3247 & TiII & 4501.2699 & TiII & 4533.9600 \\[2pt]
TiII & 4549.6216 & TiII & 4563.7575 & TiII & 4571.9713 & TiII & 4589.9466 \\[2pt]
TiII & 4657.2005 & TiII & 4779.9789 & TiII & 4805.0927 & TiII & 4911.1950 \\[2pt]
TiII & 5069.092 & TiII & 5072.2868 & TiII & 5129.1563 & TiII & 5154.0682 \\[2pt]
TiII & 5211.5305 & TiII & 5226.5385 & TiII & 5336.7860 &  &  \\[2pt]
\midrule
CrI & 3991.1135 & CrI & 4048.7797 & CrI & 4204.4650 & CrI & 4254.3517 \\[2pt]
CrI & 4274.8117 & CrI & 4296.6130 & CrI & 4359.6246 & CrI & 4362.9525 \\[2pt]
CrI & 4379.7763 & CrI & 4496.8518 & CrI & 4544.0164 & CrI & 4545.3310 \\[2pt]
CrI & 4545.9527 & CrI & 4580.0477 & CrI & 4600.7485 & CrI & 4613.3573 \\[2pt]
CrI & 4646.1620 & CrI & 4752.0875 & CrI & 4756.118 & CrI & 4764.2940 \\[2pt]
CrI & 4767.2662 & CrI & 4767.8553 & CrI & 4792.512 & CrI & 5206.0229 \\[2pt]
CrI & 5208.4094 & CrI & 5297.3762 & CrI & 5298.2715 & CrI & 5328.3238 \\[2pt]
CrI & 5345.7959 & CrI & 5409.7834 &  &  &  &  \\[2pt]
\midrule
CrII & 4242.36595 & CrII & 4558.64422 & CrII & 4588.19836 & CrII & 4592.05241 \\[2pt]
CrII & 4616.62424 & CrII & 4618.80671 & CrII & 4634.07286 & CrII & 4824.13069 \\[2pt]
CrII & 4848.24690 & CrII & 4876.39731 & CrII & 5237.32185 & CrII & 5313.5808 \\[2pt]
\midrule
MnI & 3806.72 & MnI & 4030.76 & MnI & 4033.07 & MnI & 4034.49 \\[2pt]
MnI & 4041.36 & MnI & 4461.08 & MnI & 4462.02 & MnI & 4464.68 \\[2pt]
MnI & 4472.79 & MnI & 4754.04 & MnI & 4762.38 & MnI & 4765.86 \\[2pt]
MnI & 4766.43 & MnI & 4783.42 & MnI & 4823.52 &  &  \\[2pt]
\midrule
FeI & 3694.0058 & FeI & 3695.051 & FeI & 3697.4252 & FeI & 3701.0860 \\[2pt]
FeI & 3703.6906 & FeI & 3704.4601 & FeI & 3705.5658 & FeI & 3707.8218 \\[2pt]
FeI & 3707.9196 & FeI & 3709.2461 & FeI & 3711.4080 & FeI & 3716.4419 \\[2pt]
FeI & 3719.9345 & FeI & 3721.5025 & FeI & 3722.5627 & FeI & 3724.3767 \\[2pt]
FeI & 3726.9263 & FeI & 3727.0922 & FeI & 3727.6188 & FeI & 3727.8089 \\[2pt]
FeI & 3730.3861 & FeI & 3732.3961 & FeI & 3733.3173 & FeI & 3734.8636 \\[2pt]
FeI & 3735.3235 & FeI & 3737.1313 & FeI & 3738.3049 & FeI & 3740.2392 \\[2pt]
FeI & 3742.6163 & FeI & 3743.3619 & FeI & 3743.4680 & FeI & 3744.102 \\[2pt]
FeI & 3745.5610 & FeI & 3745.8993 & FeI & 3746.9267 & FeI & 3748.2619 \\[2pt]
FeI & 3748.9643 & FeI & 3749.4851 & FeI & 3753.6107 & FeI & 3758.2327 \\[2pt]
FeI & 3760.0494 & FeI & 3763.7888 & FeI & 3765.5387 & FeI & 3767.1915 \\[2pt]
FeI & 3778.5088 & FeI & 3785.9481 & FeI & 3786.1867 & FeI & 3786.6766 \\[2pt]
FeI & 3787.8799 & FeI & 3790.0927 & FeI & 3793.4804 & FeI & 3795.0019 \\[2pt]
FeI & 3797.5146 & FeI & 3798.5111 & FeI & 3799.5473 & FeI & 3801.9837 \\[2pt]
FeI & 3805.342 & FeI & 3806.2164 & FeI & 3806.6954 & FeI & 3807.5366 \\[2pt]
FeI & 3810.7554 & FeI & 3812.9643 & FeI & 3814.5229 & FeI & 3815.8400 \\[2pt]
FeI & 3817.6392 & FeI & 3820.4249 & FeI & 3821.1776 & FeI & 3824.4435 \\[2pt]
FeI & 3825.8809 & FeI & 3827.8224 & FeI & 3834.2222 & FeI & 3836.3301 \\[2pt]
FeI & 3839.2556 & FeI & 3840.4372 & FeI & 3841.0477 & FeI & 3843.2565 \\[2pt]
FeI & 3846.800 & FeI & 3849.9664 & FeI & 3850.8176 & FeI & 3856.3713 \\[2pt]
FeI & 3859.2123 & FeI & 3859.9111 & FeI & 3865.5228 & FeI & 3867.2157 \\[2pt]
FeI & 3871.7477 & FeI & 3872.5009 & FeI & 3873.7603 & FeI & 3876.0397 \\[2pt]
FeI & 3878.0180 & FeI & 3878.5730 & FeI & 3878.6707 & FeI & 3878.7256 \\[2pt]
FeI & 3883.2797 & FeI & 3886.2820 & FeI & 3887.0480 & FeI & 3888.5132 \\[2pt]
FeI & 3888.8213 & FeI & 3891.9262 & FeI & 3893.3901 & FeI & 3894.0117 \\[2pt]
FeI & 3895.6562 & FeI & 3897.8896 & FeI & 3898.0086 & FeI & 3899.7071 \\[2pt]
FeI & 3900.515 & FeI & 3902.9455 & FeI & 3903.8976 & FeI & 3906.4795 \\[2pt]
FeI & 3906.7468 & FeI & 3916.7307 & FeI & 3917.1807 & FeI & 3918.6414 \\[2pt]
FeI & 3920.2578 & FeI & 3922.9115 & FeI & 3925.9410 & FeI & 3926.0129 \\[2pt]
FeI & 3927.920 & FeI & 3928.0827 & FeI & 3930.2964 & FeI & 3935.8122 \\[2pt]
FeI & 3940.8773 & FeI & 3941.2750 & FeI & 3942.4397 & FeI & 3946.9946 \\[2pt]
FeI & 3948.0970 & FeI & 3948.7746 & FeI & 3951.1629 & FeI & 3955.3410 \\[2pt]
FeI & 3956.4551 & FeI & 3956.677 & FeI & 3957.0181 & FeI & 3963.1002 \\[2pt]
FeI & 3966.0614 & FeI & 3967.4204 & FeI & 3967.9611 & FeI & 3969.2570 \\[2pt]
FeI & 3971.3224 & FeI & 3976.613 & FeI & 3981.7708 & FeI & 3985.3870 \\[2pt]
FeI & 3990.3731 & FeI & 3997.3919 & FeI & 3998.0525 & FeI & 4001.6615 \\[2pt]
FeI & 4005.2417 & FeI & 4006.3105 & FeI & 4009.7125 & FeI & 4012.1488 \\[2pt]
FeI & 4014.5306 & FeI & 4017.1483 & FeI & 4018.2673 & FeI & 4021.866 \\[2pt]
FeI & 4024.7248 & FeI & 4030.4883 & FeI & 4032.6272 & FeI & 4040.6379 \\[2pt]
FeI & 4043.8964 & FeI & 4045.5936 & FeI & 4045.8122 & FeI & 4057.3433 \\[2pt]
FeI & 4058.217 & FeI & 4058.7537 & FeI & 4062.4406 & FeI & 4063.5939 \\[2pt]
FeI & 4067.2709 & FeI & 4067.9775 & FeI & 4070.7705 & FeI & 4071.7377 \\[2pt]
FeI & 4073.7620 & FeI & 4076.6288 & FeI & 4084.4913 & FeI & 4085.3028 \\[2pt]
FeI & 4098.1755 & FeI & 4107.4881 & FeI & 4109.8015 & FeI & 4118.5447 \\[2pt]
FeI & 4122.5152 & FeI & 4126.1824 & FeI & 4127.6075 & FeI & 4132.0579 \\[2pt]
FeI & 4134.6773 & FeI & 4136.9974 & FeI & 4143.4143 & FeI & 4143.8678 \\[2pt]
FeI & 4147.6687 & FeI & 4149.3648 & FeI & 4153.8994 & FeI & 4154.4984 \\[2pt]
FeI & 4154.8052 & FeI & 4156.7985 & FeI & 4157.7798 & FeI & 4158.7922 \\[2pt]
FeI & 4161.0766 & FeI & 4161.4842 & FeI & 4172.1219 & FeI & 4174.9128 \\[2pt]
FeI & 4175.6358 & FeI & 4181.7544 & FeI & 4184.8915 & FeI & 4187.0387 \\[2pt]
FeI & 4187.7952 & FeI & 4191.4304 & FeI & 4195.3288 & FeI & 4196.2081 \\[2pt]
FeI & 4198.2466 & FeI & 4198.3041 & FeI & 4198.6338 & FeI & 4199.095 \\[2pt]
FeI & 4200.9239 & FeI & 4202.0289 & FeI & 4205.5382 & FeI & 4210.3433 \\[2pt]
FeI & 4217.5453 & FeI & 4219.3601 & FeI & 4222.2128 & FeI & 4224.1714 \\[2pt]
FeI & 4225.4540 & FeI & 4227.4263 & FeI & 4233.6025 & FeI & 4235.9367 \\[2pt]
FeI & 4238.8097 & FeI & 4247.4253 & FeI & 4250.1192 & FeI & 4250.7866 \\[2pt]
FeI & 4258.3156 & FeI & 4260.4741 & FeI & 4271.1534 & FeI & 4271.7602 \\[2pt]
FeI & 4278.2311 & FeI & 4282.4026 & FeI & 4288.1455 & FeI & 4294.1245 \\[2pt]
FeI & 4299.235 & FeI & 4307.9020 & FeI & 4315.0843 & FeI & 4325.7616 \\[2pt]
FeI & 4337.0460 & FeI & 4358.4989 & FeI & 4367.9033 & FeI & 4369.7715 \\[2pt]
FeI & 4383.5447 & FeI & 4390.9502 & FeI & 4404.7501 & FeI & 4407.7089 \\[2pt]
FeI & 4408.4132 & FeI & 4415.1222 & FeI & 4422.5678 & FeI & 4427.3096 \\[2pt]
FeI & 4430.1888 & FeI & 4430.6137 & FeI & 4433.2184 & FeI & 4433.7621 \\[2pt]
FeI & 4447.1301 & FeI & 4447.7170 & FeI & 4459.1173 & FeI & 4461.6525 \\[2pt]
FeI & 4461.970 & FeI & 4464.7662 & FeI & 4466.5515 & FeI & 4476.0183 \\[2pt]
FeI & 4490.0837 & FeI & 4494.5629 & FeI & 4517.5242 & FeI & 4525.1365 \\[2pt]
FeI & 4528.6139 & FeI & 4531.1478 & FeI & 4565.6616 & FeI & 4566.5142 \\[2pt]
FeI & 4580.5771 & FeI & 4541.3138 & FeI & 4592.6508 & FeI & 4602.9407 \\[2pt]
FeI & 4607.647 & FeI & 4607.647 & FeI & 4611.2786 & FeI & 4613.2025 \\[2pt]
FeI & 4625.0450 & FeI & 4632.9114 & FeI & 4637.5030 & FeI & 4638.010 \\[2pt]
FeI & 4643.4631 & FeI & 4647.4339 & FeI & 4654.4980 & FeI & 4667.4528 \\[2pt]
FeI & 4673.1633 & FeI & 4678.8455 & FeI & 4682.1108 & FeI & 4682.5601 \\[2pt]
FeI & 4685.0243 & FeI & 4691.4114 & FeI & 4694.8590 & FeI & 4694.919 \\[2pt]
FeI & 4707.2742 & FeI & 4709.0878 & FeI & 4727.3943 & FeI & 4728.5454 \\[2pt]
FeI & 4733.5913 & FeI & 4736.7731 & FeI & 4772.8027 & FeI & 4786.8067 \\[2pt]
FeI & 4788.7566 & FeI & 4789.6505 & FeI & 4798.2646 & FeI & 4799.4058 \\[2pt]
FeI & 4800.6487 & FeI & 4835.8676 & FeI & 4859.7411 & FeI & 4871.3179 \\[2pt]
FeI & 4872.1375 & FeI & 4878.2108 & FeI & 4890.7548 & FeI & 4891.4921 \\[2pt]
FeI & 4903.3099 & FeI & 4918.9937 & FeI & 4920.5028 & FeI & 4938.8135 \\[2pt]
FeI & 4939.6864 & FeI & 4942.4589 & FeI & 4946.3878 & FeI & 4957.2983 \\[2pt]
FeI & 4957.5965 & FeI & 4966.0886 & FeI & 4968.6975 & FeI & 4969.9173 \\[2pt]
FeI & 4970.4955 & FeI & 4973.1016 & FeI & 4982.4996 & FeI & 4983.2504 \\[2pt]
FeI & 4983.8526 & FeI & 4985.2526 & FeI & 4985.5469 & FeI & 4986.2223 \\[2pt]
FeI & 4991.2680 & FeI & 4994.1292 & FeI & 5001.8633 & FeI & 5006.1188 \\[2pt]
FeI & 5012.0681 & FeI & 5041.7557 & FeI & 5051.6342 & FeI & 5065.0181 \\[2pt]
FeI & 5065.1924 & FeI & 5068.7655 & FeI & 5074.7480 & FeI & 5083.3382 \\[2pt]
FeI & 5090.7736 & FeI & 5096.9977 & FeI & 5098.5719 & FeI & 5098.6978 \\[2pt]
FeI & 5099.0768 & FeI & 5107.6407 & FeI & 5110.3585 & FeI & 5110.4128 \\[2pt]
FeI & 5115.7778 & FeI & 5123.7196 & FeI & 5125.1168 & FeI & 5133.6882 \\[2pt]
FeI & 5137.3818 & FeI & 5139.2511 & FeI & 5139.4625 & FeI & 5142.4938 \\[2pt]
FeI & 5142.5409 & FeI & 5143.7234 & FeI & 5145.0933 & FeI & 5146.306 \\[2pt]
FeI & 5148.0425 & FeI & 5148.2288 & FeI & 5150.8392 & FeI & 5154.1003 \\[2pt]
FeI & 5167.4881 & FeI & 5171.5960 & FeI & 5191.4546 & FeI & 5192.3439 \\[2pt]
FeI & 5194.9414 & FeI & 5208.5936 & FeI & 5215.1803 & FeI & 5216.2737 \\[2pt]
FeI & 5226.8620 & FeI & 5227.1891 & FeI & 5232.9400 & FeI & 5254.9551 \\[2pt]
FeI & 5263.3059 & FeI & 5266.5550 & FeI & 5269.5370 & FeI & 5270.3560 \\[2pt]
FeI & 5273.1632 & FeI & 5281.7900 & FeI & 5283.6207 & FeI & 5302.299 \\[2pt]
FeI & 5307.3607 & FeI & 5324.1787 & FeI & 5328.0383 & FeI & 5328.5313 \\[2pt]
FeI & 5332.8994 & FeI & 5339.9290 & FeI & 5341.0237 & FeI & 5349.7373 \\[2pt]
FeI & 5353.3732 & FeI & 5364.8709 & FeI & 5367.466 & FeI & 5369.9615 \\[2pt]
FeI & 5371.4893 & FeI & 5379.5737 & FeI & 5383.3688 & FeI & 5393.1673 \\[2pt]
FeI & 5397.1276 & FeI & 5404.1513 & FeI & 5405.7749 & FeI & 5410.9095 \\[2pt]
FeI & 5415.1989 & FeI & 5424.0678 & FeI & 5429.6964 & FeI & 5434.5235 \\[2pt]
FeI & 5446.9164 & FeI & 5455.6092 & FeI & 5463.2759 & FeI & 5466.3958 \\[2pt]
FeI & 5466.9874 & FeI & 5473.9001 & FeI & 5481.2427 & FeI & 5481.4383 \\[2pt]
FeI & 5487.7457 & FeI & 5497.5157 & FeI & 5501.4649 & FeI & 5506.7788 \\[2pt]
FeI & 5535.4175 & FeI & 5543.9353 & FeI & 5546.5055 & FeI & 5546.9920 \\[2pt]
FeI & 5554.8947 & FeI & 5563.5999 & FeI & 5565.7036 & FeI & 5569.6177 \\[2pt]
FeI & 5572.8420 & FeI & 5576.0884 & FeI & 5586.7555 & FeI & 5602.9447 \\[2pt]
FeI & 5615.6435 & FeI & 5624.5418 & FeI & 5633.9459 & FeI & 5638.2618 \\[2pt]
FeI & 5641.4336 & FeI & 5655.1761 & FeI & 5658.8161 & FeI & 5708.0942 \\[2pt]
FeI & 5709.3779 & FeI & 5711.8483 & FeI & 5712.1312 & FeI & 5747.9538 \\[2pt]
FeI & 5753.1223 & FeI & 5762.9918 & FeI & 5862.3561 & FeI & 5914.1117 \\[2pt]
FeI & 5914.2009 & FeI & 5930.1795 & FeI & 5952.7180 & FeI & 6020.0132 \\[2pt]
FeI & 6024.0576 & FeI & 6056.0043 & FeI & 6065.482 & FeI & 6137.6913 \\[2pt]
FeI & 6191.558 & FeI & 6252.5550 & FeI & 6254.2581 & FeI & 6256.3611 \\[2pt]
FeI & 6393.6009 & FeI & 6400.0008 & FeI & 6408.0179 & FeI & 6411.6489 \\[2pt]
FeI & 6421.3504 & FeI & 6430.8460 & FeI & 6462.725 & FeI & 6592.9134 \\[2pt]
FeI & 6593.8701 & FeI & 6633.7492 & FeI & 6663.4417 & FeI & 6677.9865 \\[2pt]
FeI & 6841.339 & FeI & 6842.6854 & FeI & 6843.6555 & FeI & 6855.1617 \\[2pt]
FeI & 6858.1493 & FeI & 7411.1539 & FeI & 7445.7504 & FeI & 7495.0670 \\[2pt]
FeI & 7511.0200 & FeI & 7531.1447 & FeI & 7586.0182 & FeI & 7445.7504 \\[2pt]
FeI & 7748.2689 & FeI & 7780.5568 & FeI & 7832.1963 & FeI & 7937.1401 \\[2pt]
FeI & 7945.8464 & FeI & 7998.9453 & FeI & 8046.0474 & FeI & 8085.1718 \\[2pt]
FeI & 8220.3784 & FeI & 8327.0558 & FeI & 8387.7719 & FeI & 8514.0716 \\[2pt]
FeI & 8688.6249 & FeI & 8790.5211 & FeI & 8793.3433 & FeI & 8824.2205 \\[2pt]
FeI & 9738.572 & FeI & 9997.5981 & FeI & 10145.563 & FeI & 10216.313 \\[2pt]
FeI & 10452.7508 & FeI & 10469.654 &  &  &  &  \\[2pt]
\midrule
FeII & 3906.037 & FeII & 4173.450 & FeII & 4178.8546 & FeII & 4233.1627 \\[2pt]
FeII & 4351.7629 & FeII & 4385.379 & FeII & 4416.818 & FeII & 4491.400 \\[2pt]
FeII & 4508.2805 & FeII & 4515.3344 & FeII & 4520.2214 & FeII & 4522.6278 \\[2pt]
FeII & 4549.197 & FeII & 4549.4665 & FeII & 4555.8881 & FeII & 4576.328 \\[2pt]
FeII & 4583.8292 & FeII & 4596.0084 & FeII & 4629.3317 & FeII & 4731.439 \\[2pt]
FeII & 4923.9216 & FeII & 5018.4354 & FeII & 5169.0282 & FeII & 5197.5678 \\[2pt]
FeII & 5234.6236 & FeII & 5275.9968 & FeII & 5316.6098 & FeII & 5316.777 \\[2pt]
FeII & 5362.7509 & FeII & 5362.9698 & FeII & 5948.4183 & FeII & 5952.5183 \\[2pt]
FeII & 5961.7058 & FeII & 5984.8484 & FeII & 5991.3721 & FeII & 6147.734 \\[2pt]
FeII & 6149.231 & FeII & 6238.375 & FeII & 6247.559 & FeII & 6383.7296 \\[2pt]
FeII & 6416.9303 & FeII & 6432.6772 & FeII & 6456.3805 & FeII & 6516.0766 \\[2pt]
FeII & 7462.414 & FeII & 7515.1029 & FeII & 7711.7219 & FeII & 9997.5980 \\[2pt]
\midrule
NiI & 3783.53 & NiI & 4605.00 & NiI & 4714.42 & NiI & 5035.37 \\[2pt]
NiI & 5080.52 & NiI & 5081.11 & NiI & 5084.08 & NiI & 5476.91 \\[2pt]
NiI & 6643.64 &  &  &  &  &  &  \\[2pt]
\midrule
ZnI & 4680.13590 & ZnI & 4722.15690 & ZnI & 4810.53210 & ZnI & 6362.3458 \\[2pt]
\midrule
SrII & 4077.714 & SrII & 4215.524 & SrII & 10327.309 &  &  \\[2pt]
\midrule
YII & 3774.330 & YII & 3788.693 & YII & 4177.528 & YII & 4374.933 \\[2pt]
YII & 4883.682 & YII & 4900.118 & YII & 5087.418 & YII & 5662.922 \\[2pt]
\midrule
ZrII & 3836.76 & ZrII & 4149.20 & ZrII & 4208.98 &  &  \\[2pt]
\midrule
BaII & 4554.033 & BaII & 4934.077 & BaII & 5853.675 & BaII & 6141.713 \\[2pt]
BaII & 6496.898 &  &  &  &  &  &  \\[2pt]
\end{longtable}

\end{document}